\documentclass{PoS}
\usepackage[authoryear,square]{natbib}
\bibpunct{(}{)}{;}{a}{}{,}

\let\ifpdf\relax

\title{Cosmic Dawn and Epoch of Reionization Foreground Removal with the SKA}

\ShortTitle{CD/EoR Foreground Removal}

\author{\speaker{Emma Chapman},$^{a}$ Anna Bonaldi,$^{b}$ Geraint Harker,$^{a}$ Vibor Jeli\'{c},$^{cd}$ Filipe B. Abdalla,$^{ae}$ Gianni Bernardi,$^{fgh}$ J\'{e}r\^{o}me Bobin,$^{i}$ Fred Dulwich,$^{j}$ Benjamin Mort,$^{j}$ Mario Santos,$^{fkl}$ and Jean-Luc Starck$^{i}$\\
\llap{$^a$}Department of Physics \& Astronomy, University College London, Gower Street, London, WC1E 6BT, U.K.\\
\llap{$^b$}Jodrell Bank Centre for Astrophysics, School of Physics \& Astronomy, University of Manchester, Oxford Road, Manchester M13 9PL, U.K. \\
\llap{$^c$} Kapteyn Astronomical Institute, University of Groningen, PO Box
800, 9700 AV Groningen, the Netherlands \\
\llap{$^d$} ASTRON - the Netherlands Institute for Radio Astronomy, PO Box
2, 7990 AA Dwingeloo, the Netherlands \\
\llap{$^e$} Department of Physics and Electronics, Rhodes University, PO Box 94, Grahamstown, 6140, South Africa \\
\llap{$^f$} SKA SA, 3rd Floor, The Park, Park Road, Pinelands, 7405, South Africa \\
\llap{$^g$} Department of Physics and Electronics, Rhodes University, Grahamstown, South Africa \\
\llap{$^h$} Harvard-Smithsonian Center for Astrophysics, Cambridge, MA, USA \\
\llap{$^i$} CEA, IRFU, Service d'Astrophysique, 91191 Gif-sur-Yvette Cedex, France \\
\llap{$^j$} Oxford e-Research Centre, University of Oxford, Keble Rd, Oxford, UK, OX1 3QG \\
\llap{$^k$} Physics Department, University of the Western Cape, Cape Town 7535, South Africa \\
\llap{$^l$} CENTRA, Instituto Superior Tecnico, Universidade de Lisboa, Portugal \\

E-mail: \email{eow@star.ucl.ac.uk}, \email{anna.bonaldi@manchester.ac.uk}, \email{g.harker@ucl.ac.uk}, \email{vjelic@astro.rug.nl}, \email{fba@star.ucl.ac.uk}, \email{gbernardi@ska.ac.za}, \email{jbobin@cea.fr}, \email{fred.dulwich@oerc.ox.ac.uk}, \email{benjamin.mort@oerc.ox.ac.uk}, \email{mgrsantos@uwc.ac.za}, \email{jstarck@cea.fr}}

\abstract{The exceptional sensitivity of the SKA will allow observations of the Cosmic Dawn and Epoch of Reionization (CD/EoR) in unprecedented detail, both spectrally and spatially. This wealth of information is buried under Galactic and
extragalactic foregrounds, which must be removed accurately and precisely in order to reveal the cosmological signal. This problem has been addressed already for the previous generation of radio telescopes, but the application to SKA is different in many aspects.

In this chapter we summarise the contributions to the
field of foreground removal in the context of high redshift and high sensitivity 21-cm measurements. We use a state-of-the-art simulation of the SKA Phase 1
observations complete with cosmological signal, foregrounds and
frequency-dependent instrumental effects to test both parametric and
non-parametric foreground removal methods. We compare the recovered
cosmological signal using several different statistics and explore one
of the most exciting possibilities with the SKA --- imaging of the
ionized bubbles. 

We find that with current methods it is possible to remove the foregrounds with great accuracy and to get impressive power spectra and images of the cosmological signal. The frequency-dependent PSF of the instrument complicates this recovery, so we resort to splitting the observation bandwidth into smaller segments, each of a common resolution. 

If the foregrounds are allowed a random variation from the smooth power law along the line of sight, methods exploiting the smoothness of foregrounds or a parametrization of their behaviour are challenged much more than non-parametric ones. However, we show that correction techniques can be implemented to restore the performances of parametric approaches, as long as the first-order approximation of a power law stands.
}

\FullConference{
Advancing Astrophysics with the Square Kilometre Array\\
June 8-13, 2014\\
Giardini Naxos, Italy}

\begin{document}

\section{Introduction}

The statistical detection of the 21-cm reionization signal depends on an accurate and robust method for removing the foregrounds from the total signal. The first attempts to address this problem focused on exploiting the angular fluctuations of
the 21-cm signal, e.g. \citet{dimatteo02,oh03,dimatteo04}, but the 21-cm
signal was found to be swamped by various foregrounds. The focus then
moved on to the frequency correlation of the foregrounds, with the
cross-correlation of pairs of maps used as a cleaning step
\citep{zal04,santos05}. This naturally evolved into methods which exploited the correlation of the foreground across whole segments of, or the entire bandwidth of, the observation: line of sight (LOS) fitting. This LOS fitting (to be discussed in much greater detail in Sec.~\ref{sec:methods}) has been numerically shown to be the optimal method for power spectrum recovery \citep{liu11}.

For this Chapter, we choose to concentrate on comparing the major LOS methods in the field, alongside the much more recent idea of avoiding the foregrounds altogether \citep[e.g.,][]{dillon13} by focusing analysis on the area of Fourier space where the foregrounds are sub-dominant to the cosmological signal (see Sec.~\ref{sec:win}). 

One of the `saving graces' of foreground contamination is its smoothness over frequency space. While the foregrounds are expected to be highly correlated on the order of MHz, the cosmological signal in comparison is expected to be highly uncorrelated, e.g. \citet{dimatteo02,gnedin04}. LOS methods can be divided into subcategories of parametric and non-parametric methods. Both aim to find the form of the smooth foreground function along frequency for each line of sight and subtract this from the total signal leaving residuals of noise, fitting errors and the cosmological signal.

The majority of early line of sight methods in the literature can be termed
parametric as at some point they assume a specific form for the
foregrounds, for example a polynomial (see Sec.~\ref{sec:para}). Despite the successes of the parametric methods, the fact remains that the form of the foregrounds is not definitively known across the frequency range and resolution of interest. Too great an assumption of their spectral form risks introducing a large element of uncertainty into the cosmological signal detection. It is with this argument that, more recently, `blind' methods have been investigated. These allow the data to determine the form of the foregrounds, without assuming any particular shape beforehand (see Sec.~\ref{sec:nonpara}). This has obvious advantages for a cosmological era so far not directly observed, but results are often not as promising as parametric results when applied to simulations. Arguably this is common sense, since in parametric methods one has modelled the foregrounds based on the simulation knowledge. If these methods were applied to foregrounds of different shape to the accepted form, they would risk suffering a large drop in accuracy. 

Though there are now multiple proof-of-principle papers relating to LOS fitting for the EoR signal recovery, there has been little consideration given to which method aids the recovery of CD/EoR information most efficiently, accurately or precisely. Though the comparison of polynomial-fitting methods to any new method introduced is fairly common, the comparison between more complicated parametric, non-parametric methods and indeed foreground avoidance, is rare, with the exception of brief comparisons in \citet{GU13,patil14,chapman14}. In this Chapter we aim to compare non-parametric and parametric methods on a SKA Phase 1 CD/EoR simulation, comparing the recovered signals both in terms of statistics and imaging. 

In Sec.~\ref{sec:methods} we introduce the five methods used in this chapter to mitigate the simulated foreground contamination. In Sec.~\ref{sec:sims} we describe the state-of-the-art SKA Phase 1 simulation, consisting of cosmological signal, foregrounds and instrumental noise, used in this chapter. We show the results of applying the methods to these simulations in Sec.~\ref{sec:results} before making our conclusions in Sec.~\ref{sec:conc}. 

\section{Comparing Foreground Removal Methods}
\label{sec:methods}
In this section we describe the methods of LOS foreground removal applied in this chapter. We divide the methods into those which assume a functional form for the foreground signal (parametric) and those which loosen the constraints on this form somewhat (non-parametric).

\subsection{Parametric Methods}
\label{sec:para}
\subsubsection{Polynomial fitting}\label{sec:poly}
The simplest method for foreground removal in total intensity is polynomial fitting
in frequency or log frequency, e.g. \citet{mcquinn06, morales06, gleser08, jelic2008, bowman09, liu09, petrovic11}. 

The usual method of polynomial fitting is to fit the total observed spectrum along the line of sight with a smooth function such as a $n$-th order polynomial: $\log T_{b,fg}(\nu) = a_0 + \sum^n_{i=1}a^i \log \nu^i$. The order of polynomial varies slightly between different papers, for example \citet{wang06} set $n=2$ while \citet{jelic2008} sets $n=3$. 

One should be careful in choosing the order of the polynomial to 
perform the fitting. If the order of the polynomial is too small, the foregrounds will be under-fitted and the EoR signal could be dominated and corrupted by the fitting residuals. If the order of the polynomial is too big, the EoR signal could be fitted out. For this work we will follow \cite{jelic2008} and perform the fitting in log space using a 3rd order polynomial.

\subsubsection{CCA (Correlated Component Analysis)}
\label{subsubsec:CCA}
In this section we describe the main principles of operation of the Fourier-domain Correlated Component Analysis (CCA) method. More details on the method and on its application to the HI signal can be found in \citet{ricciardi2010} and \citet{bonaldibrown}, respectively. 
The CCA is a ``model learning'' algorithm, which estimates the frequency spectrum of the foreground components from the data exploiting second-order statistics.  
This method was developed for the Cosmic Microwave Background (CMB); its ability to improve the modelling of the poorly known anomalous microwave emission has been demonstrated in \citet{bonaldi2007}, \citet{special} and \citet{gouldbelt}. 

We start by modelling the data in the $uv$ plane as a linear mixture of the foreground components. For each point in the $uv$ plane we write
\begin{equation}
\vec{x}=\mathbf{B}\mathbf{A}\vec{s}+\vec{n}\label{modhcca}.
\end{equation}
The vectors $\vec{x}$ and $\vec{n}$ contain the data and the noise in Fourier space, respectively; the vector $\vec{s}$ contains the astrophysical foregrounds; the diagonal matrix $\mathbf{B}$ contains the instrumental dirty beams in Fourier space and the  matrix $\mathbf{A}$, called the mixing matrix, contains the intensity of the foreground components at all frequencies. The 21-cm signal is modelled as a noise term, contributing to $\vec{n}$ together with the instrumental noise.

The additional assumptions made by the CCA are that the mixing matrix is constant within the considered area of the sky, and that its unknown elements can be reduced by adopting a suitable parametrization $\mathbf{A}=\mathbf{A}(\vec{p})$. For example, in the following, a power law is assumed for the synchrotron component with unknown, spatially constant, spectral index. For the free-free, we adopt a power-law behaviour with fixed spectral index of -2.08. When necessary, we can adopt other parametric models having more degrees of freedom. Though CCA allows the data to estimate the parameters of the model, it does exploit a parametrization, and therefore it is classified as a parametric method.  

Once we have an estimate of the mixing matrix, using a relation between the cross-spectra of the data, we can invert eq.~(\ref{modhcca}) and reconstruct the foreground components as $\vec{ \hat s}=\mathbf{W}\vec{x}$ directly in the Fourier domain.

The cleaning of the HI signal consists of subtracting the estimated foreground components $\vec{ \hat s}$ at all frequencies. We perform the subtraction as:
\begin{equation}
\vec{x}-\mathbf{R}\mathbf{\hat H}\vec{\hat s}
\end{equation}
where $\mathbf{R}$ is a diagonal matrix whose elements are chosen to improve the subtraction by minimizing the power of the residuals at each frequency. This step compensates for small errors in the parametric model adopted by the CCA, which result in a slight over/underestimation of the amplitude of the foregrounds at a given frequency. The effectiveness of this approach is tested with the simulation R2 (see Sections 3.4 and 4.4). It is important to note that this minimization approach could be applied to all methods, and as such is a way of mitigating the weaknesses of the parametric method as opposed to the inherent ability for the non-parametric method to deal with foregrounds differing from our models.

\subsection{Non-Parametric Methods}
\label{sec:nonpara}
\subsubsection{Wp smoothing}\label{subsubsec:wp}

Wp smoothing is a technique, introduced to 21-cm work by \citet{NONPAR_09}, to fit the foregrounds LOS-by-LOS. The aim is to directly exploit the physical expectation that the foregrounds are smooth, so in this sense the foreground separation is not completely blind. It does not, however, assume a specific parametric form for the foregrounds, or anything about their spatial structure. Wp penalises changes in curvature, with roughness measured `apart from inflection points (IPs)', hence the name `Wendepunkt' (Wp), the German word for `inflection point'.

In the $i$-th LOS, we have a set $\{(\nu^i_1,s^i_1),(\nu^i_2,s^i_2),\ldots,(\nu^i_n,s^i_n)\}$ of observations in $n$ frequency channels, which we wish to fit with a smooth function $f(\nu)$. Since Wp smoothing always applies to one LOS, from now on we will drop the superscript for clarity. Wp smoothing takes as its measure of roughness the integrated change of curvature. If $\kappa$ is the radius of curvature, the standardized change of curvature is $\kappa '/\kappa \approx f'''/f''$, where the approximation, which we adopt here, holds exactly at local extrema ($f'=0$) and becomes singular at IPs ($f''=0$).

We therefore separate out the IPs, writing $f''$ as a smooth function $g(\nu)$ multiplied by a polynomial with the IPs as its roots. With the IPs specified, we find $f$ by performing a penalised fit to the data, where the penalty term is given by a measure of the integrated change in curvature of $g$ multiplied by a smoothing parameter, $\lambda$.

This formulation of Wp smoothing is given by \citet{MAC93,MAC95}, who derived a boundary value problem, the solution of which is the function $f$ we seek. Different algorithms have been proposed to solve this system \citep{MAC89,GU13}, but we use that outlined by \citet{NONPAR_09}. Unfortunately, these methods currently take $\sim 1\,\mathrm{s}$ to solve for a single sightline, depending on the value of $\lambda$, making Wp smoothing relatively slow for large data cubes.

In principle, the choice of a value for $\lambda$ should be determined
by the level of smoothness we expect in our foregrounds. In the limit
of $\lambda\to 0$, $f$ becomes the best-fitting function with the
given inflection points, while for $\lambda\to\infty$ it becomes the
best-fitting polynomial of degree $n_w+2$. Here, we fix $n_w=0$ and
choose $\lambda$ based on the performance of the method in
simulations. The quality of the fit is quite insensitive to the value
of $\lambda$ up to at least a factor of 2, however.

\subsubsection{GMCA}

\begin{figure*}
\begin{centering}
\includegraphics[trim=0cm 7cm 0cm 7cm,clip=true,width=65mm]{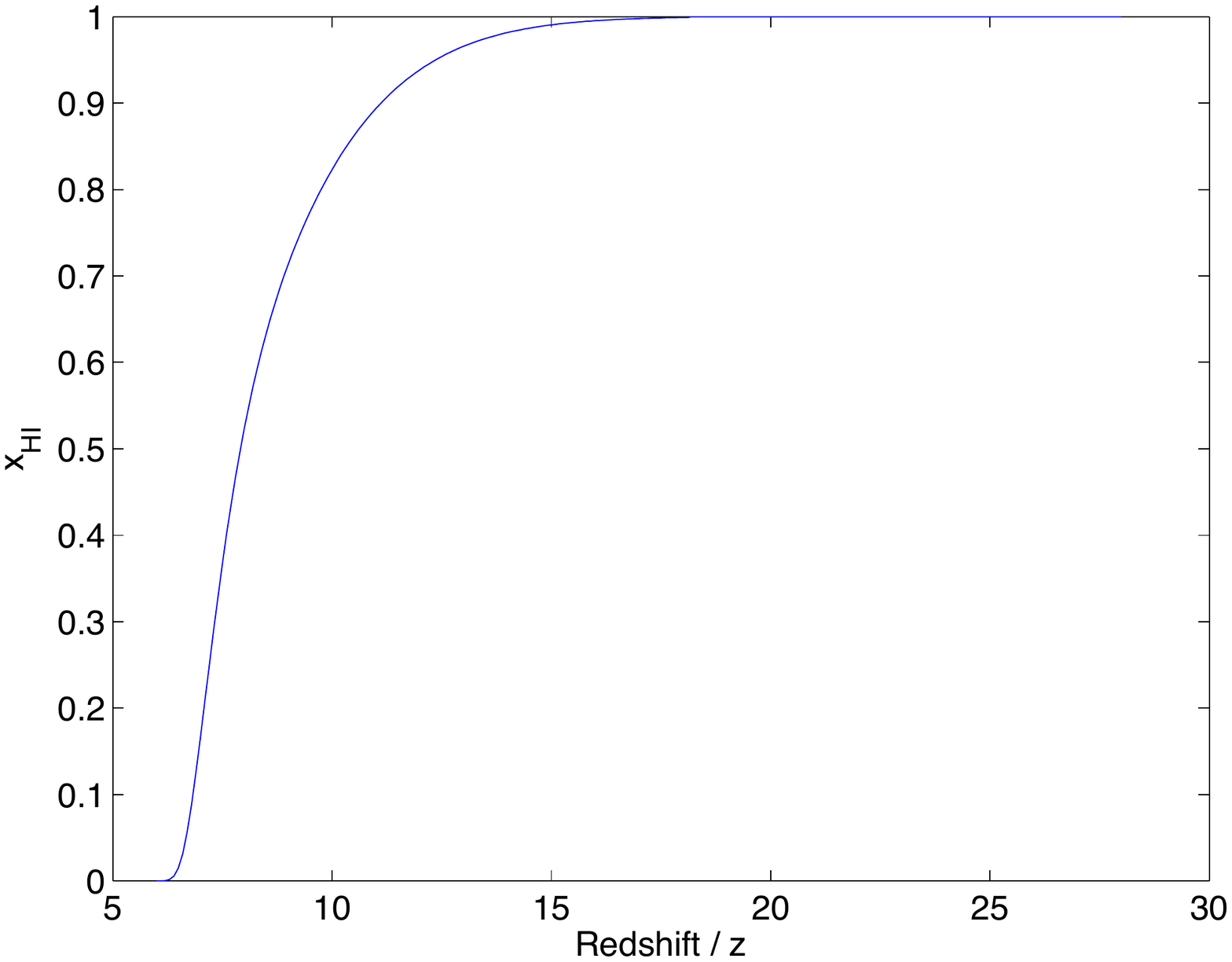} 
\includegraphics[trim=0cm 7cm 0cm 7cm,clip=true,width=65mm]{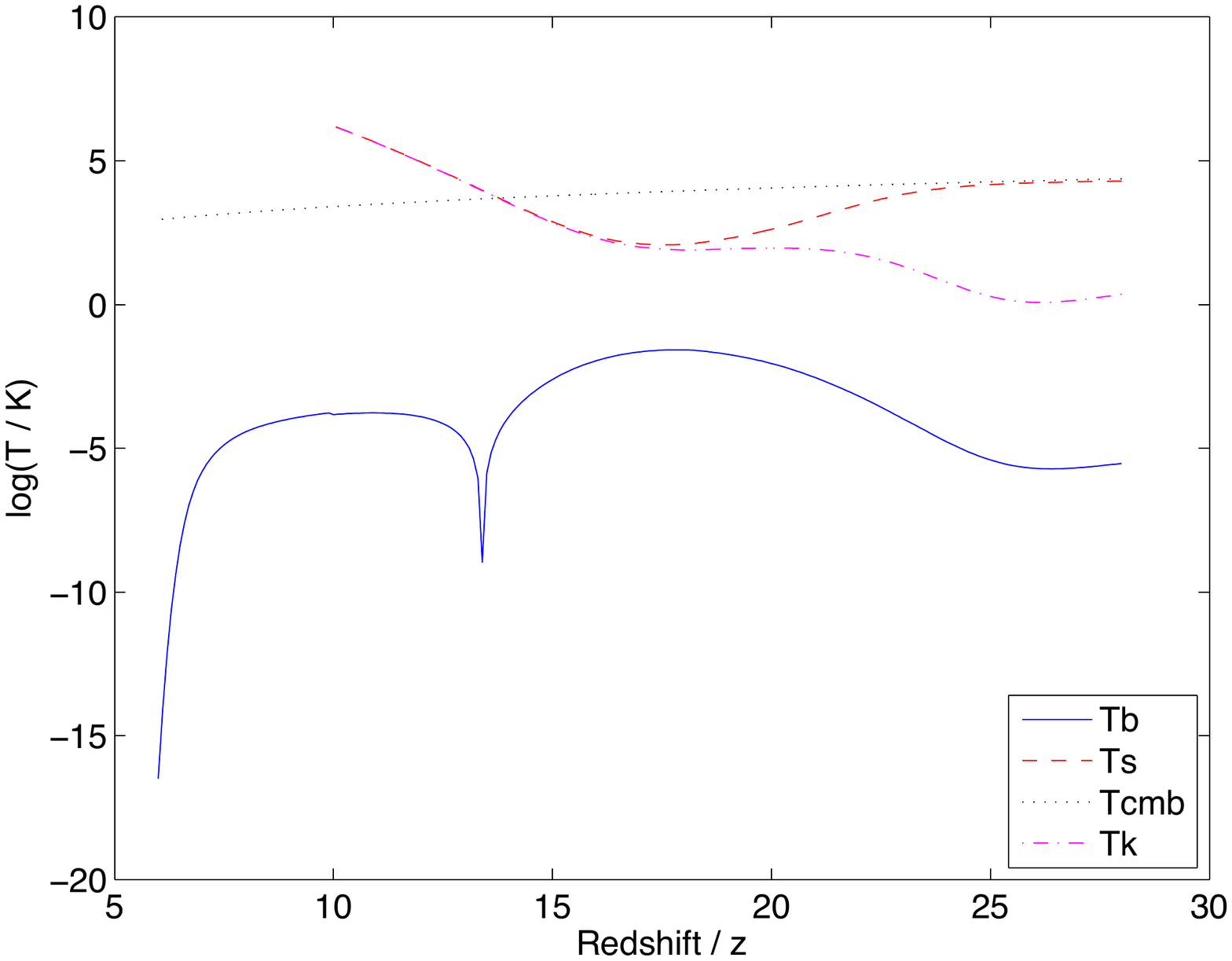} 
\caption{Left: The evolution of neutral hydrogen fraction with redshift. Right: The evolution of $T_s,T_{cmb},T_K$ and $T_b$ with redshift. The absolute log is taken of $T_b$.}
\label{fig:cssim}
\end{centering}
\end{figure*}

Blind source separation (BSS) uses a mixing model $\vec{x}=\mathbf{A}\vec{s}+\vec{n}$, where $\vec{x}$ is the observed data $\mathbf{A}$ is the mixing matrix, $\vec{s}$ is the unmixed data components and $\vec{n}$ is the noise. What defines a BSS problem is the need to estimate both $\mathbf{A}$ and $\vec{s}$ with no prior knowledge of either (note the difference to CCA, where a prior form for $\mathbf{A}$ was assumed). Methods differ in their approach to this estimation with, for example, the independent component analysis technique FastICA (\citet{hyvarinen99},\citet{hyvarinen01} and applied to EoR data by \citet{chapman12}) assuming statistical independence of the components $\vec{s}$. Here we utilise another BSS technique, Generalized Morphological Component Analysis (GMCA), which assumes morphological diversity and sparsity of the foregrounds in order to model them. This approach originated with \citet{zibulevsky01} who suggested that one could find a basis set in which the components to be found would be sparsely represented, i.e. a basis set where only a few of the coefficients would be non-zero. With the components being unlikely to have the same few non-zero coefficients one could then use this sparsity to more easily separate the mixture. 
We attempt to recover the cosmological signal as a residual of the process, i.e. it is actually part of $\vec{n}$. We can expand the data $\vec{x}$ in a wavelet basis and seek an unmixing scheme, through the estimation of $\mathbf{A}$,
which yields the sparsest components $\vec{s}$ in the wavelet domain. 

For more technical details about GMCA, we refer the interested reader to \citet{bobin07,bobin08a,bobin08b,bobin12}, where it is shown that sparsity, as used in GMCA, allows for a more precise estimation of the mixing matrix $\mathbf{A}$ and more robustness to noise than ICA-based techniques such as FastICA. For a previous application of GMCA to EoR data see \citet{chapman13}.

This component separation method has been applied to the Planck PR1 data to estimate a low-foreground CMB map \citep{GMCA_PR1}. In this context, sparsity is well adapted to remove naturally non-Gaussian and heterogeneous components such as foregrounds.

\begin{figure*}[ht]
\begin{centering}
\includegraphics[width=140mm]{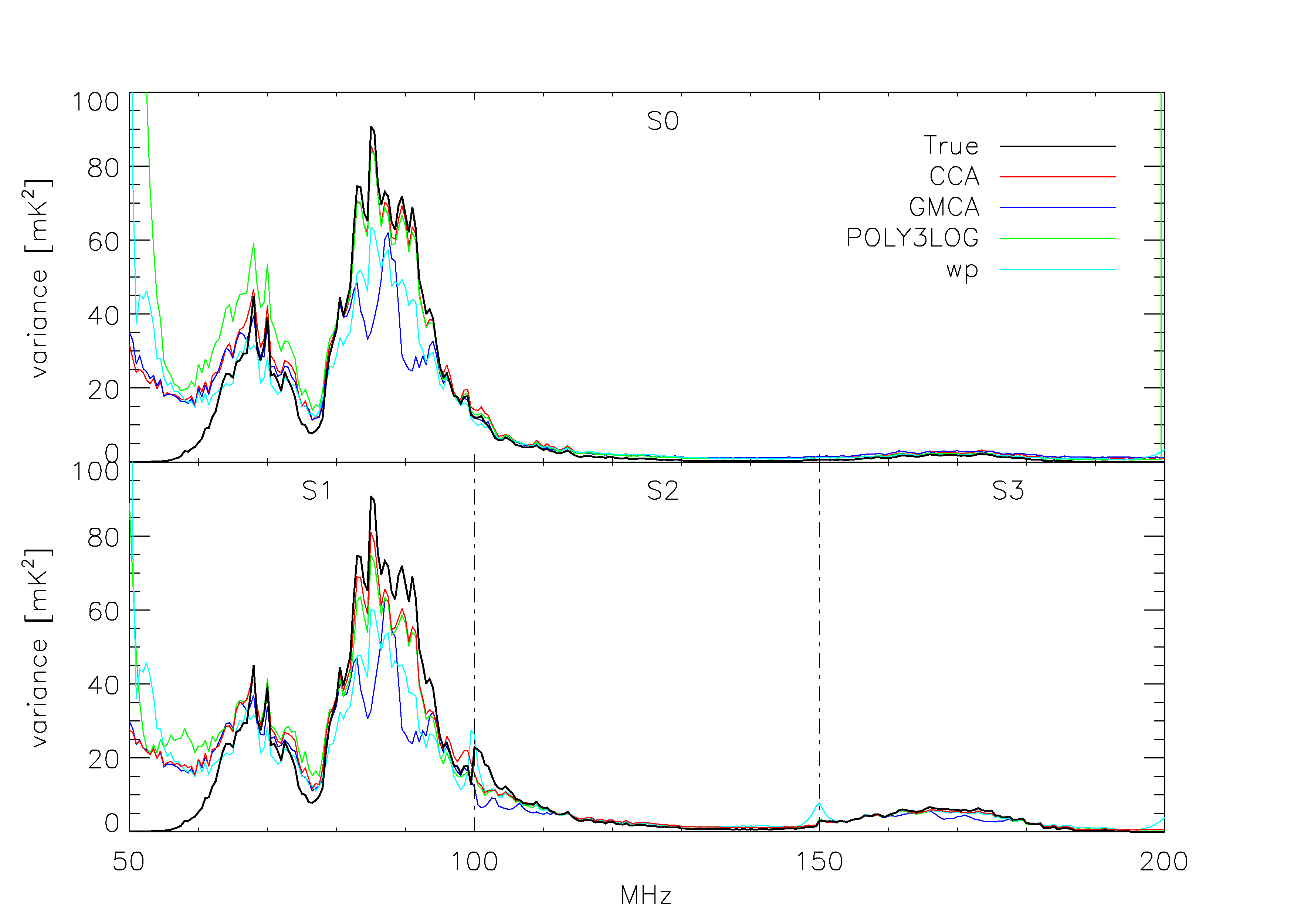}
\caption{Variance of the simulated cosmological signal (black line) and of the reconstructed cosmological signals for the 4 foreground removal methods (coloured lines) for the S0 cube (top) and the S1, S2 and S3 cubes (bottom).}
\label{fig:var}
\end{centering}
\end{figure*}

\begin{figure*}[!ht]
\begin{centering}
\includegraphics[width=120mm]{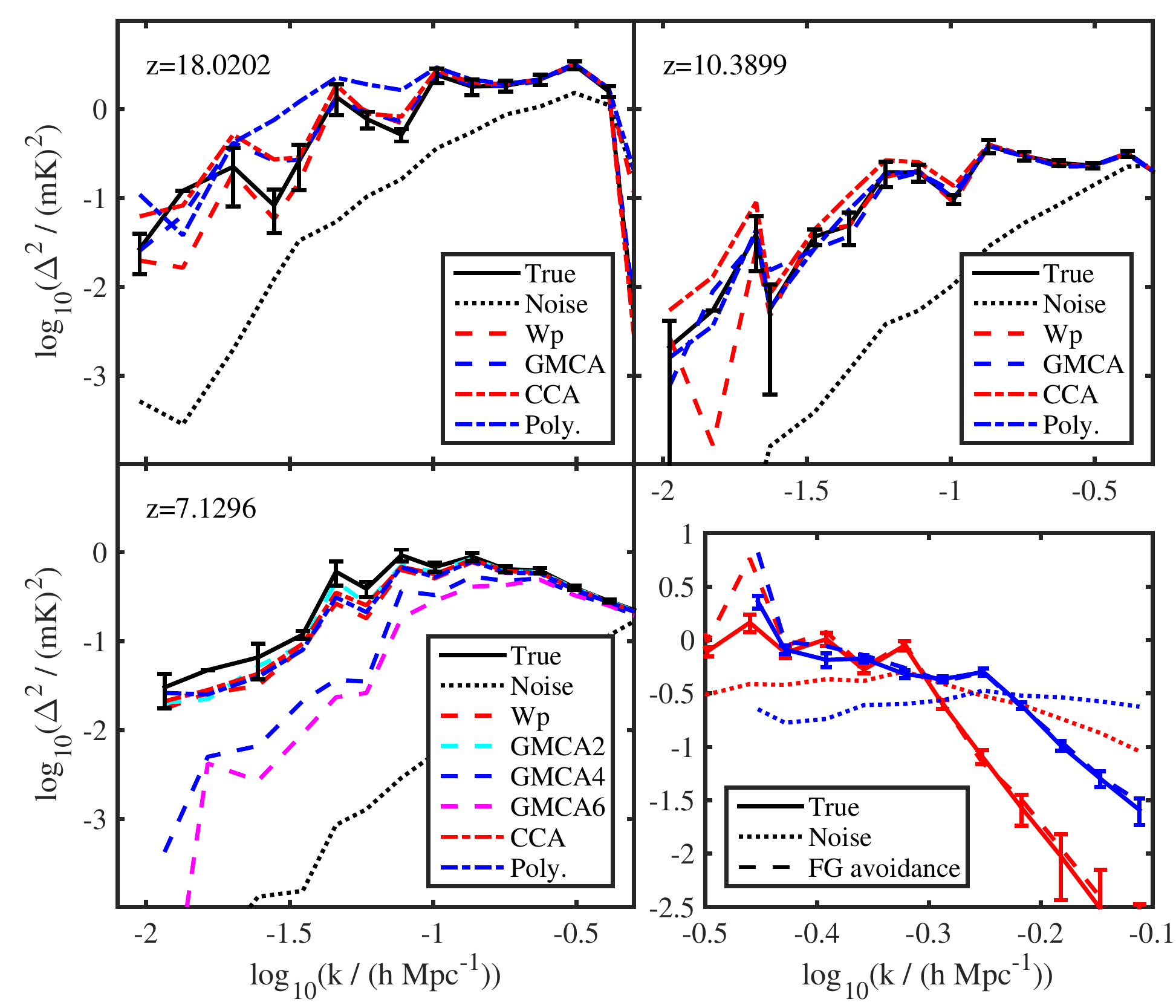}
\caption{Spherically averaged power spectrum of the simulated cosmological signal, reconstructed cosmological signal, and noise. The top left, top right and bottom left panels show, respectively, power spectra computed in an $8\,\mathrm{MHz}$ slice in the centre of S1, S2 and S3. The central redshift of each slice is shown in the panel. Different line styles and colours show the true (input) signal, the noise, and the reconstructed signal for the four different foreground removal methods, as described in the legend. The axes are the same in each of these panels. Note that we show two extra GMCA lines according to foreground models with 2 and 6 components in the bottom left panel. We do not show the highest $k$ points since they are strongly affected by the point spread function. The bottom right panel shows the application of foreground avoidance. Here, the different line styles show the true signal, the noise and the recovered signal, with the different colours used for the data at different redshifts (red for S2, blue for S3), with cutoffs in $k_{\rm los}$ as described in the text. The result for S1 is poorer, and is not shown in order to avoid compressing the scale of the plot; note that the scale in this panel is different from the other three. In all cases, the power spectra are computed after applying a Hanning taper in the frequency direction to avoid ringing: this is particularly important for foreground avoidance, since it mitigates the aliasing of unsubtracted foreground power to high $k_{\rm los}$.}
\label{fig:3D_ps}
\end{centering}
\end{figure*}

\begin{figure*}[ht]
\begin{centering}
\includegraphics[width=70mm]{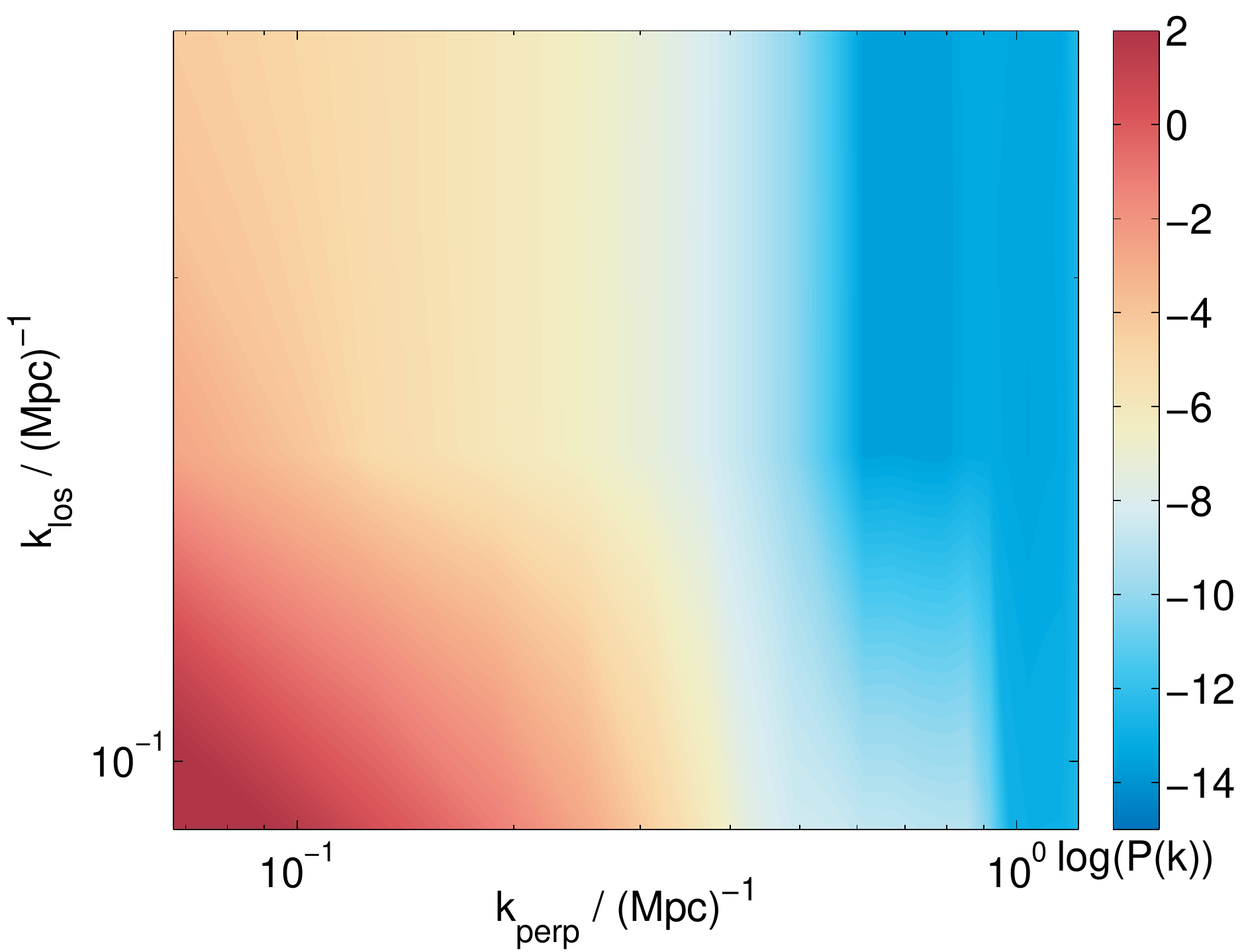}
\includegraphics[width=70mm]{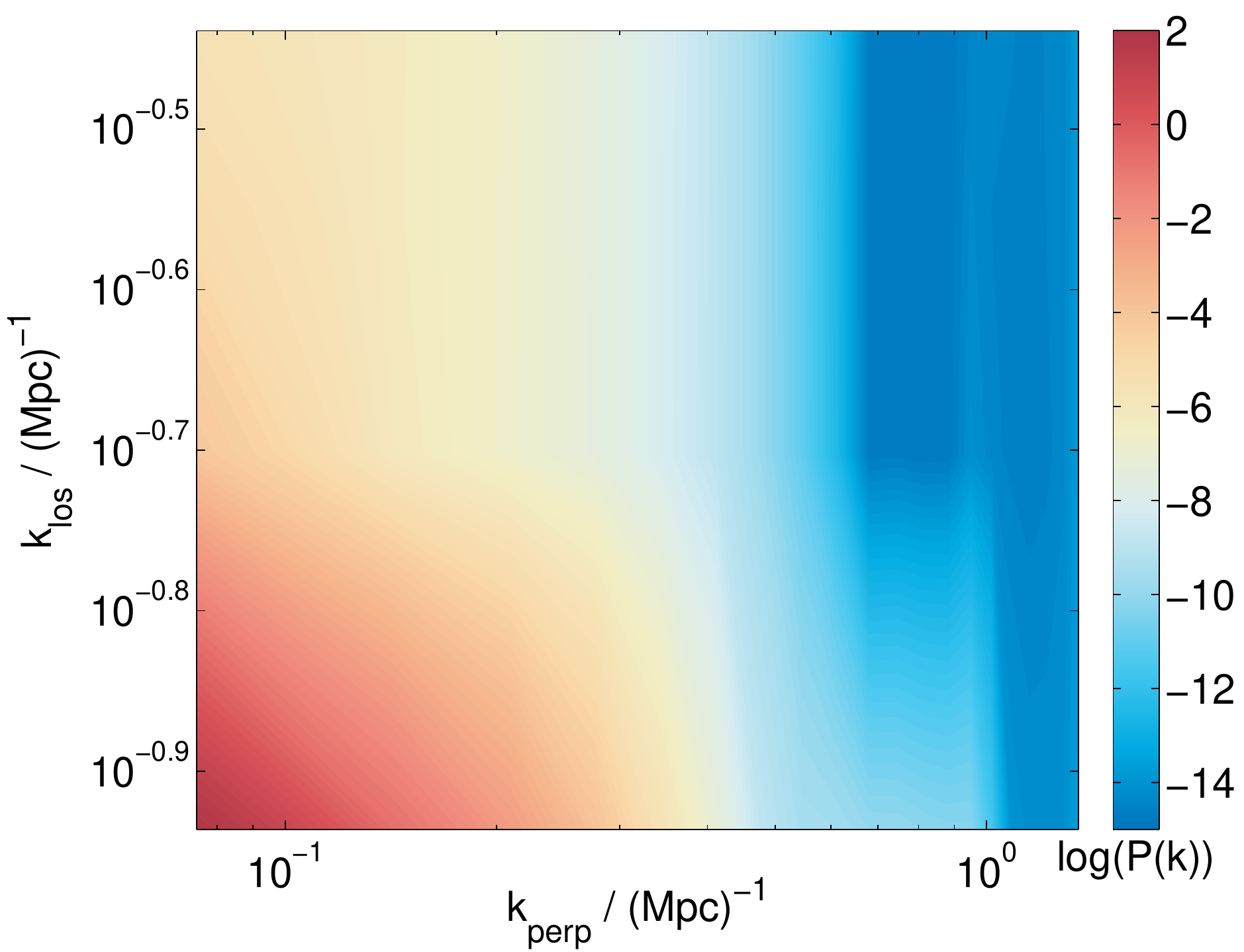}
\includegraphics[width=70mm]{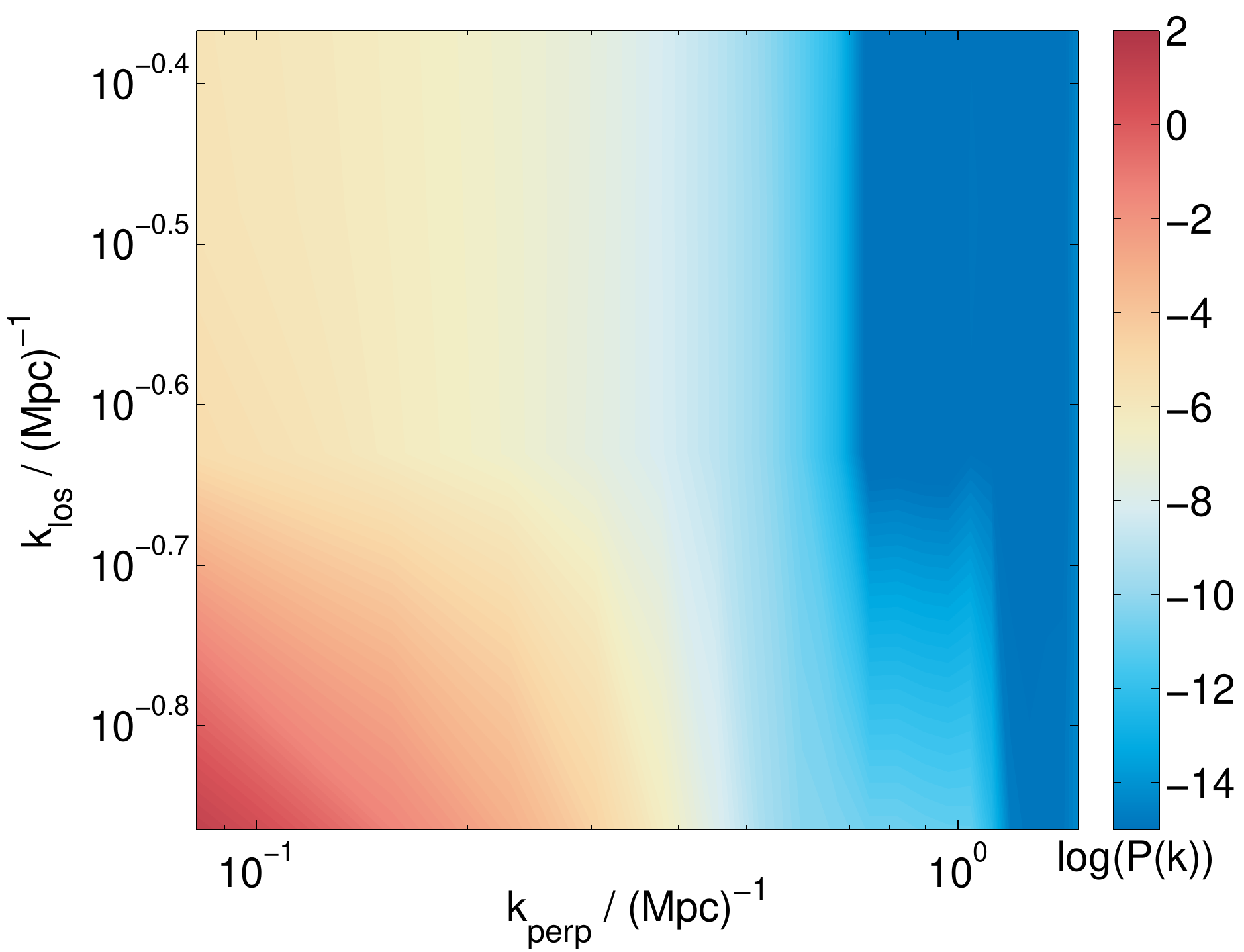}
\caption{Cylindrical power spectrum of the S0 cube at 75 MHz, 125 MHz and 175 MHz (in reading order). The foreground contamination can be clearly seen for $k_{\rm los}<10^{-0.93}$, $k_{\rm los}<10^{-0.7}$ and $k_{\rm los}<10^{-0.63}\,\mathrm{Mpc}^{-1}$ respectively, along with the action of the PSF at high $k_{\rm perp}$.}
\label{fig:win}
\end{centering}
\end{figure*}

\subsection{Foreground Avoidance}
\label{sec:win}
While the methods so far introduced have been focused on removing the foreground from the total signal, there has recently been discussion of avoiding the foregrounds instead. The coupling between the expected frequency smoothness of foregrounds and the unavoidable frequency dependent response of an interferometer leaves a characteristic footprint in $k$-space, separating an area over which foreground dominates, from a region which is virtually foreground free --- a so-called `EoR window'. The boundaries of this EoR window have been discussed at length in the literature \citep{datta10,vedantham12,morales12,trott12,parsons12,liu14a,LIU14b} as well as seen in early results from low frequency observations \citep{bernardi13,dillon13,pober13}. The largest extent of the EoR window happens to be at low $k_{\rm perp}$ where the frequency dependent response of the instrument is smooth, excluding the low $k_{\rm los}$ mode that are likely to be dominated by foregrounds. When constructing other statistics such as a spherical power spectrum, one can then simply ignore all the foreground-dominated modes and use the $k$ modes inside the EoR window alone. While this will result also in the loss of any cosmological signal outside the EoR window, the remaining region should be free of foreground contamination, if the foregrounds are indeed so well-defined even in the face of instrumental calibration errors. In comparison, while foreground removal allows the possibility of recovering the cosmological signal in all modes, there is also the possibility of foreground fitting bias being introduced on all modes.

\section{SKA Phase 1 Simulations}
\label{sec:sims}
\subsection{Cosmological Signal}
We simulate the cosmological signal using the semi-numerical reionization code \textsc{simfast21} \citep{santos10}.

We create initial conditions boxes on a grid of $1024^3$ cells before producing ionization and brightness temperature boxes on a grid of $256^3$. The boxes are a constant $1\,\mathrm{Gpc}$ in side length and are output between redshifts 6 and 28 at separations of $\mathrm{d}z=0.1$. As the aim of this chapter is to assess the effectiveness of foreground removal methods, as opposed to studying a particular model of reionization, we have simply used the default options for \textsc{simfast21}, for example a minimum halo mass of 1.0e8 $M_{\odot}$. When calculating the brightness temperature, $T_b$, it is usual to assume that $T_s >> T_{cmb}$ where $T_s$ is the spin temperature and $T_{cmb}$ is the CMB temperature. However, this assumption breaks down at high redshift due to the increasing effect of the gas temperature ($T_K$) and Lyman-alpha coupling on the spin temperature. We calculate the full spin temperature for all redshifts above 10. We plot the evolution of the neutral hydrogen fraction, $x_{HI}$, and $T_s,T_{cmb},T_K$ and $T_b$ in Fig. \ref{fig:cssim}. 

From the real space $T_b$ boxes we create an observation cube, or `light cone', with the LOS axis evolving in redshift and a constant 5 degree field of view. 

\subsection{Foregrounds}

Though there have been foreground observations at frequencies relevant to LOFAR using WSRT \citep{bernardi09,bernardi2010} the foreground contamination at the frequencies and resolution of LOFAR remains poorly constrained. As a result, foreground models directly relevant for this paper rely on using constraints from observations at different frequency and resolution ranges. These constraints are used to normalize the necessary extrapolations made from observations to create a model relevant for LOFAR-EoR observations.

In general, the foreground components are modelled as power laws in 3+1 dimensions (i.e. three spatial and frequency) such that $T_\mathrm{b} \propto \nu^{\beta}$.

The foreground simulations used in this paper are obtained using the foreground models described in \citet{jelic2008,jelic10}. The foreground contributions considered in these simulations are Galactic synchrotron emission, Galactic free-free emission and extragalactic foregrounds.

\subsection{Instrumental Effects}
The instrumental effects were modelled using the OSKAR simulator\footnote{http://www.oerc.ox.ac.uk/~ska/oskar} and a list of preliminary station positions for the SKA1-LOW.

The images of the SKA-low PSF were produced by assuming full correlation between all 866 core stations, where the maximum baseline length is 5.29 km. The telescope was located at 52.7 degrees latitude. Baseline coordinates were generated for a 12-hour synthesis observation, with a phase centre on the sky at apparent equatorial coordinates $(\alpha, \delta)$ = (218, 34.5) degrees. A 5-minute sampling interval was used to give 144 snapshots each of 374545 baselines. PSF images were then generated in CASA across a 5-degree field-of-view using 256 w-projection planes. The noise is then normalized according to a 1000 hour integration time using the prescription described in e.g.\ \citet{thompson01}.

\subsection{Cubes for Analysis}
In order to simulate an observation, one normally constructs a `dirty' cube whereby the cosmological signal and foregrounds are convolved with the same frequency-dependent PSF used to construct the instrumental noise. However, the standard foreground removal methods require the channels to have common resolution in order to work optimally. In which case we use five cubes, all with channels separated by 0.5\,MHz, in our analysis:

\begin{itemize}
\item S0: A cube running from 50--200\,MHz consisting of the clean foregrounds and cosmological signal convolved with the PSF at 50\,MHz and the instrumental noise constructed with a sampling equivalent to the 50\,MHz sampling in each channel.  
\item S1: A cube running from 50--99.5\,MHz consisting of the clean foregrounds and cosmological signal convolved with the PSF at 50\,MHz and the instrumental noise constructed with a sampling equivalent to the 50\,MHz sampling in each channel.
\item S2: A cube running from 100--149.5\,MHz consisting of the clean foregrounds and cosmological signal convolved with the PSF at 100\,MHz and the instrumental noise constructed with a sampling equivalent to the 100\,MHz sampling in each channel.
\item S3: A cube running from 150--200\,MHz consisting of the clean foregrounds and cosmological signal convolved with the PSF at 150\,MHz and the instrumental noise constructed with a sampling equivalent to the 150\,MHz sampling in each channel.
\item R2: We multiply each channel of the clean foreground cube by a random number drawn from a Gaussian distribution with standard deviation of 0.05, simulating a 5\% random wiggle along the line of sight. This foreground cube is then convolved and used to construct a cube as described in S2.
\end{itemize}

The adjustment of the entire frequency range to a common resolution in S0 results in the loss of a lot of high resolution information at high frequency. There is a balance to be made between the amount of information lost and the amount of data the methods need to provide an optimal foreground estimate. We therefore test S1, S2 and S3 in order to assess whether the methods are able to make good signal recoveries of the higher resolution information at higher frequencies, despite a third of the data being available to constrain the foregrounds.

We construct R2 (R - rough) in order to test the reliance of the methods on the smoothness of the foregrounds. We expect methods with strong constraints on the smoothness such as polynomial, CCA and, to some degree, Wp smoothing to be more affected than GMCA which places no explicit prior on the smoothness. This roughness could be interpreted as an inherent roughness of the foregrounds themselves or as a simple approximation of an instrumental calibration error such a leakage of polarized foregrounds.

\section{Results}
\label{sec:results}
\subsection{Variance}

We computed the variance of the true input cosmological signal and of the reconstructed cosmological signals for each of the 4 foreground removal methods. The variance has been computed for a pixel size of 2.3\,arcmins. Given the low noise, the results are stable for changes of the pixel size. The results are shown in Fig.~\ref{fig:var} for both the S0 cube and the collated S1, S2 and S3 cubes. 

All methods show an excess variance at $\nu <60\,$MHz which is due to foreground residuals. The results are good for the other frequencies. GMCA and Wp smoothing somewhat underestimate the variance in the 80--90\,MHz frequency range. For all methods the variance is correctly recovered over a broad frequency range.

There is not an apparent major disadvantage to any of the methods by splitting the cube into three segments and so we pursue analysis of only the S1, S2 and S3 cubes in the following, in order to retain as much spatial information as possible while conforming to the common resolution requirements of the methods. 

One might wonder at the two peaks in variance below 100 MHz. While the clean signal does not show such clear peaks, the action of the realistic SKA PSF on the clean cosmological signal induces this effect.

\begin{figure*}
\begin{centering}
\includegraphics[trim=0cm 0cm 0cm 2.9cm,clip=true,width=65mm]{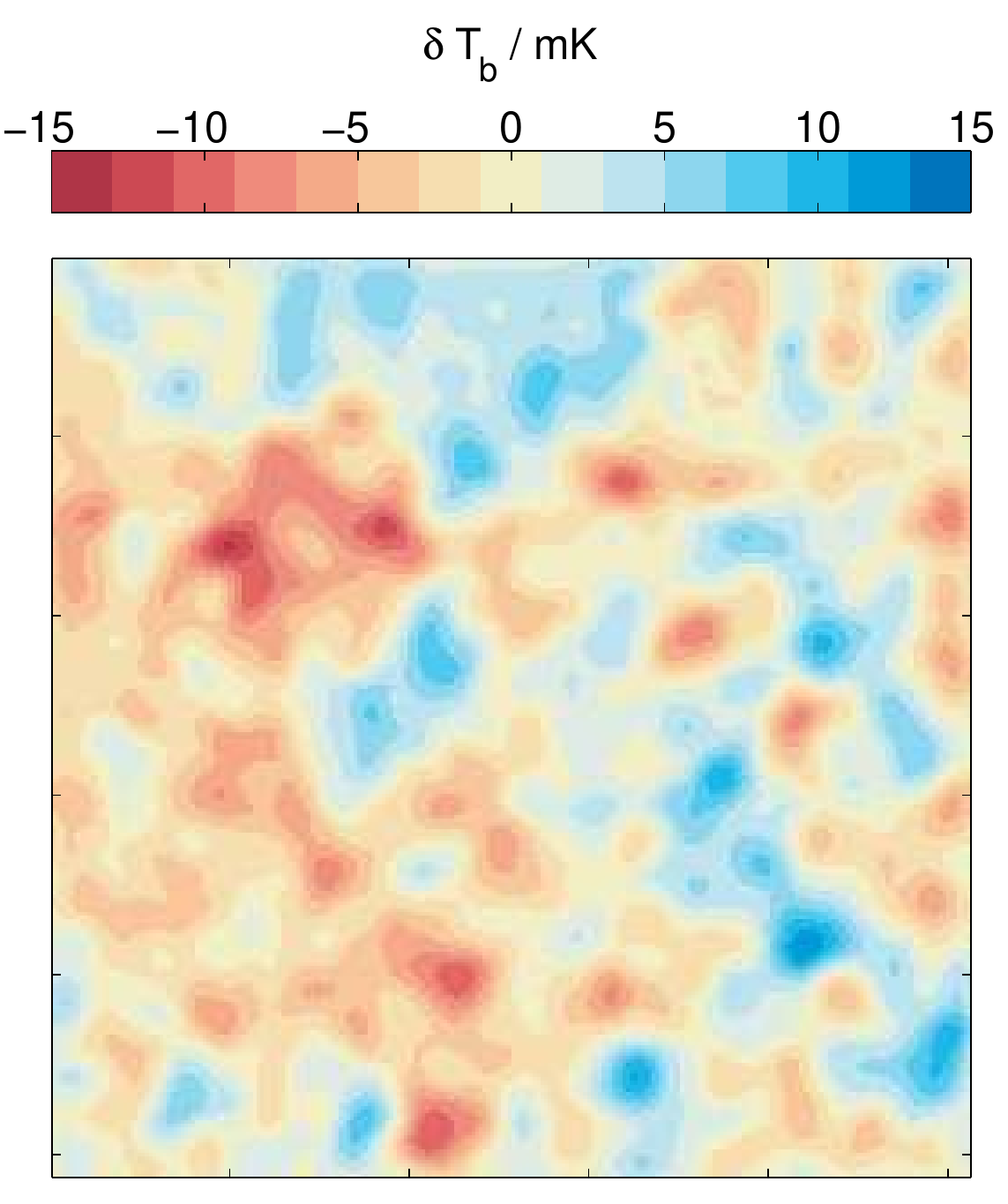} 
\includegraphics[trim=0cm 0cm 0cm 2.9cm,clip=true,width=65mm]{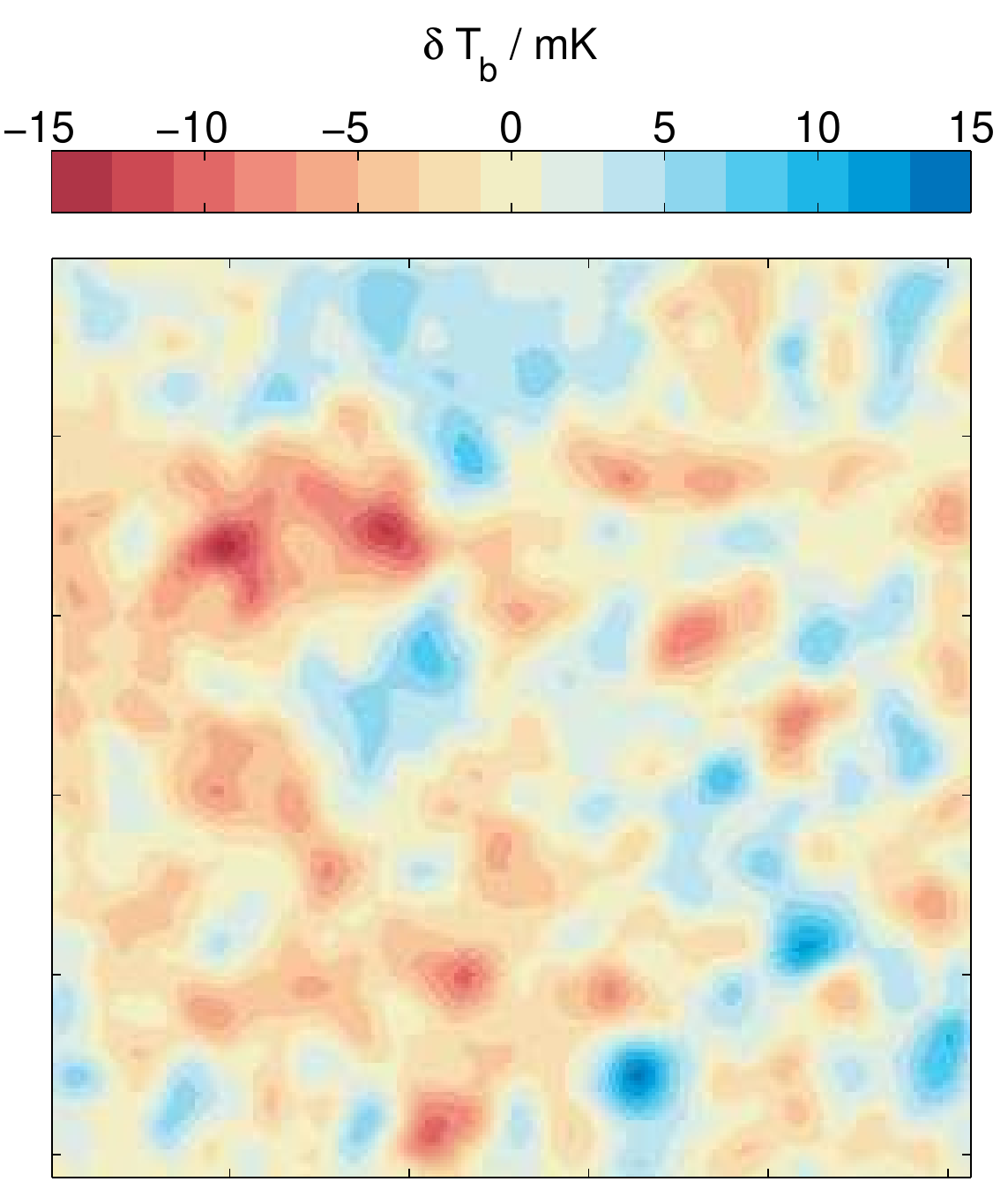} 
\includegraphics[trim=0cm 0cm 0cm 2.9cm,clip=true,width=65mm]{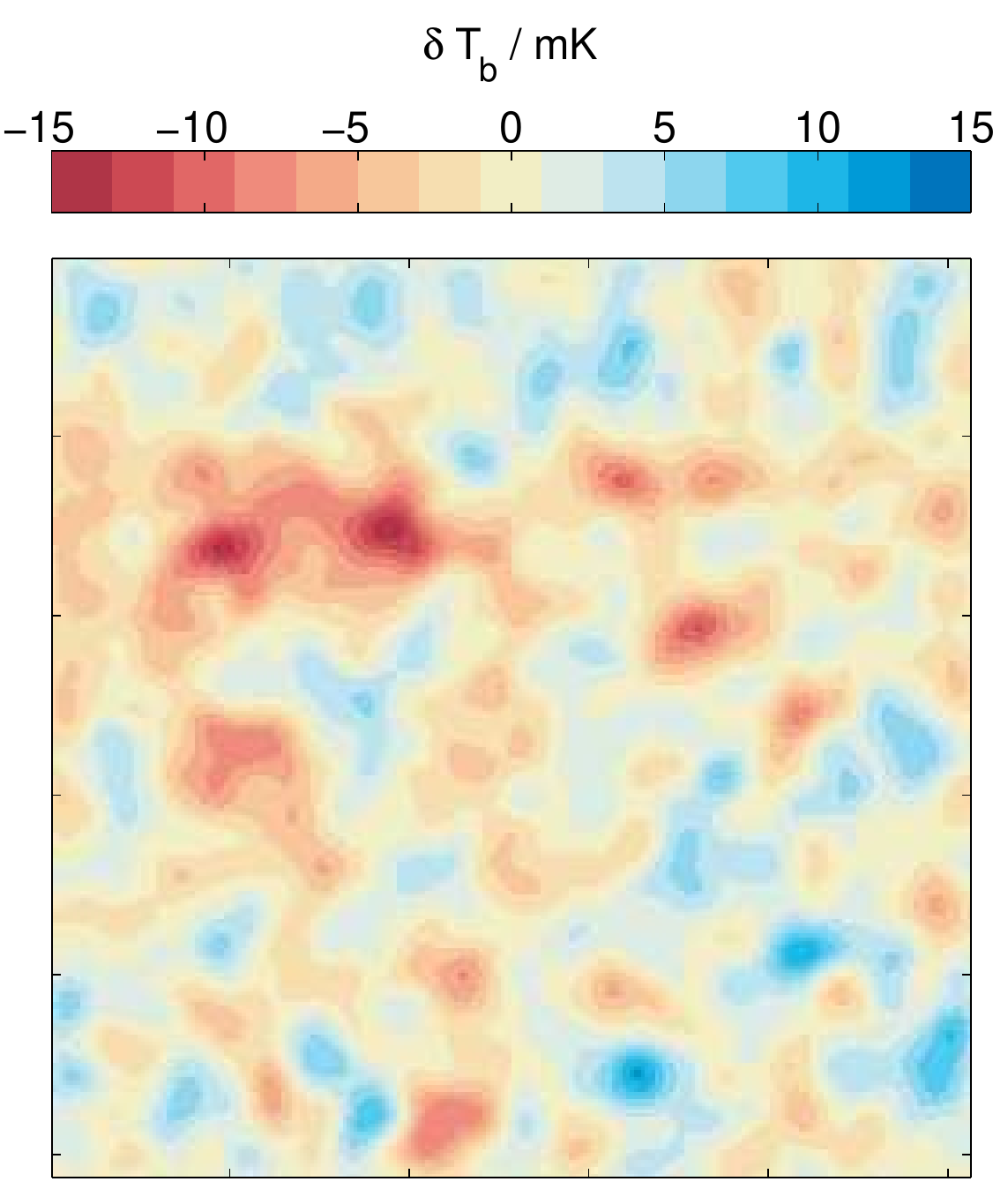} 
\includegraphics[trim=0cm 0cm 0cm 2.9cm,clip=true,width=65mm]{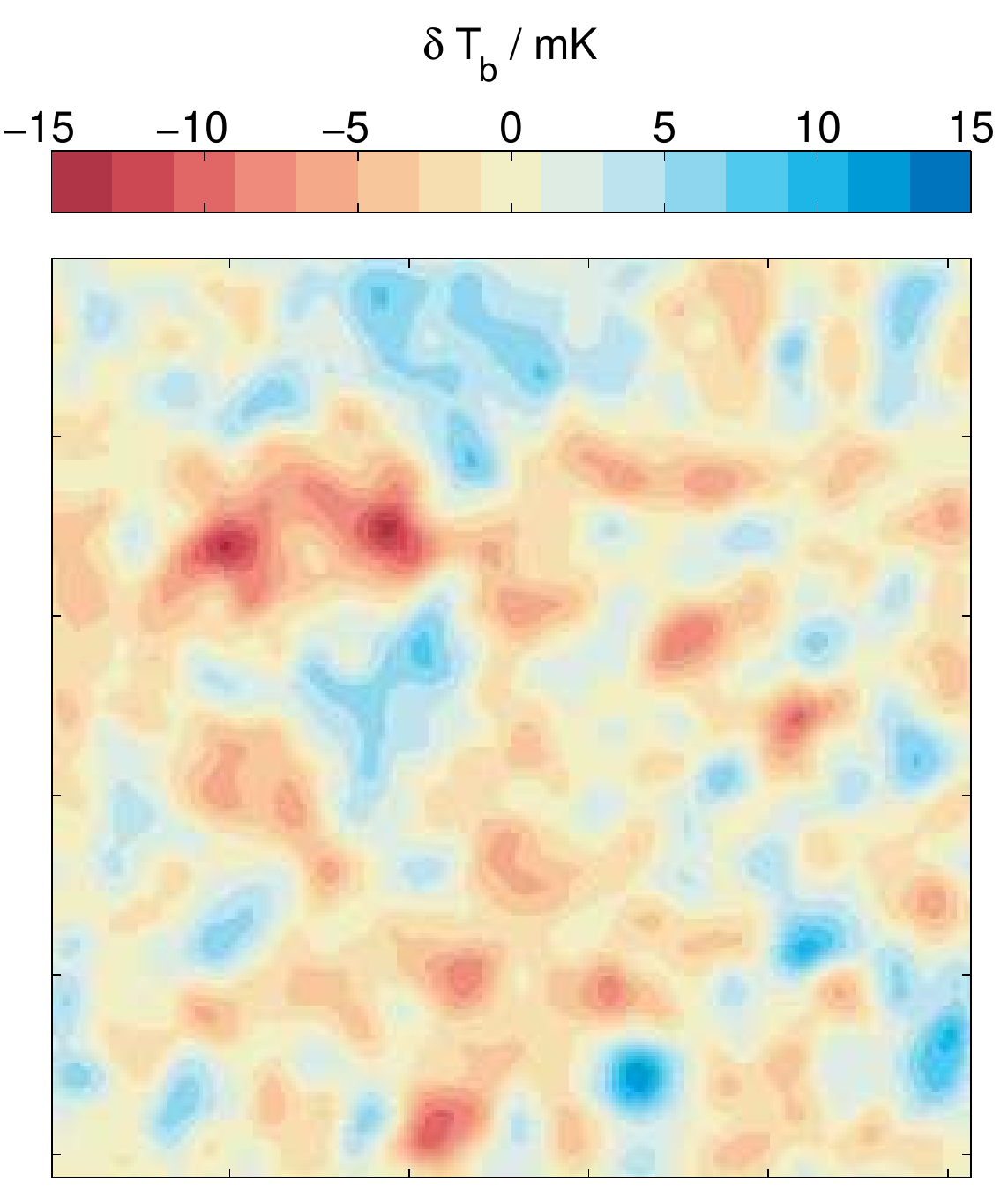}
\includegraphics[width=65mm]{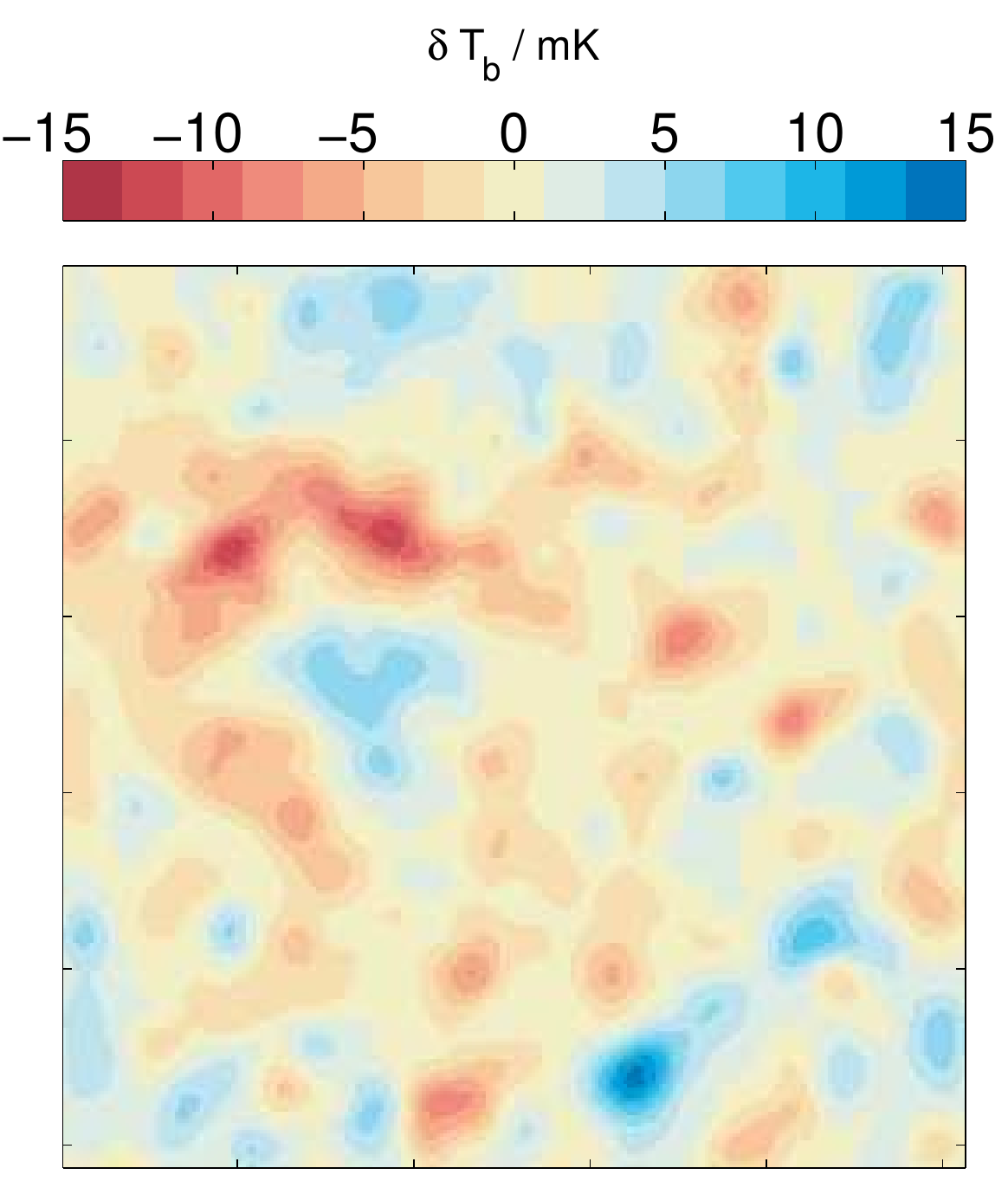}
\includegraphics[width=68mm]{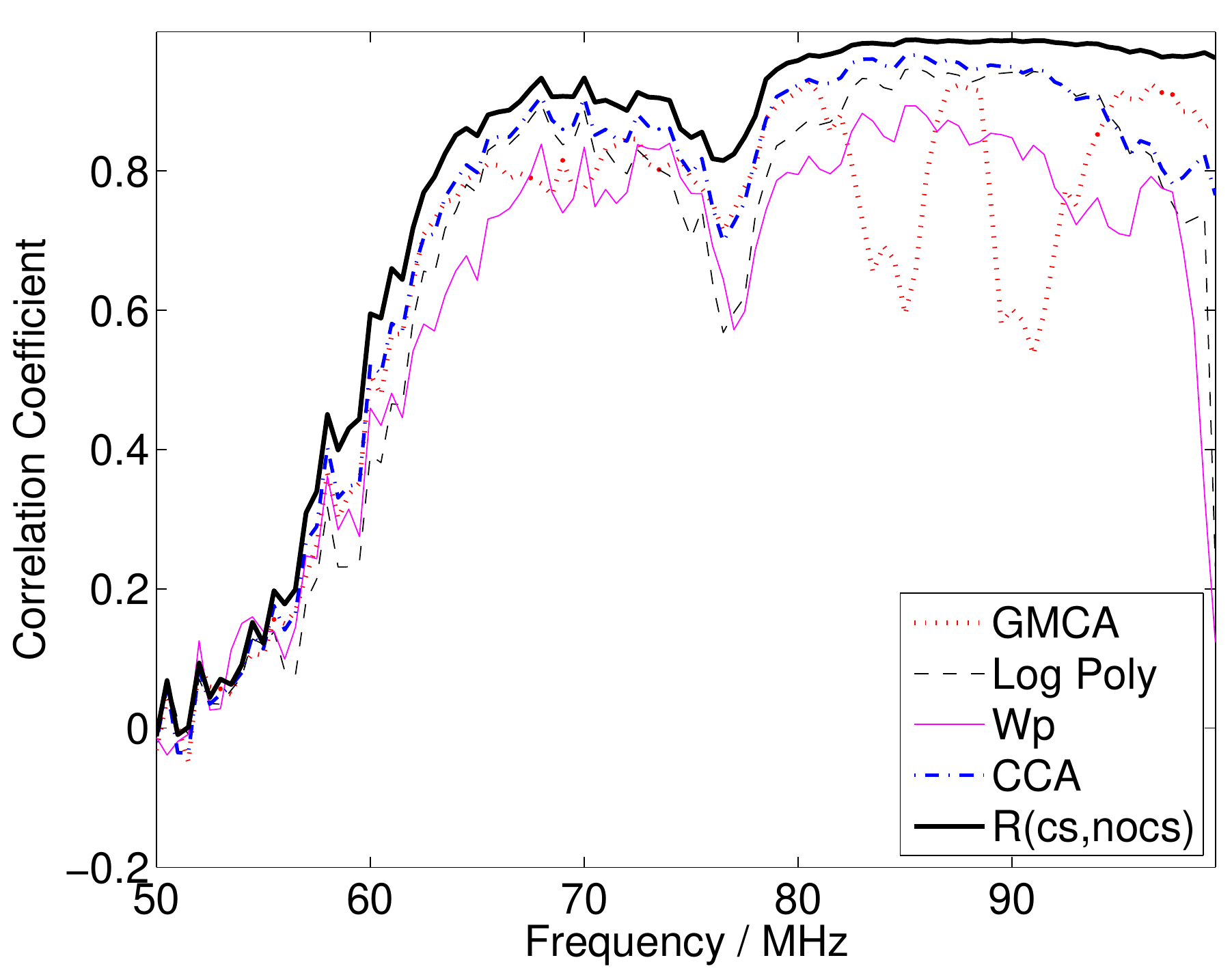} 
\caption{Top two rows, reading order: the smoothed residual maps of S1 at 75\,MHz from the Polynomial, CCA, Wp and GMCA methods. Bottom row, left to right: The smoothed cosmological signal at 75\,MHz and the correlation coefficient relating to residuals vs. cosmological signal. The best theoretical recovery possible, whereby foreground fitting errors are zero, is shown by the correlation between noise plus simulated cosmological signal vs. simulated cosmological signal in solid black.}
\label{fig:im_50}
\end{centering}
\end{figure*}

\begin{figure*}
\begin{centering}
\includegraphics[trim=0cm 0cm 0cm 2.9cm,clip=true,width=65mm]{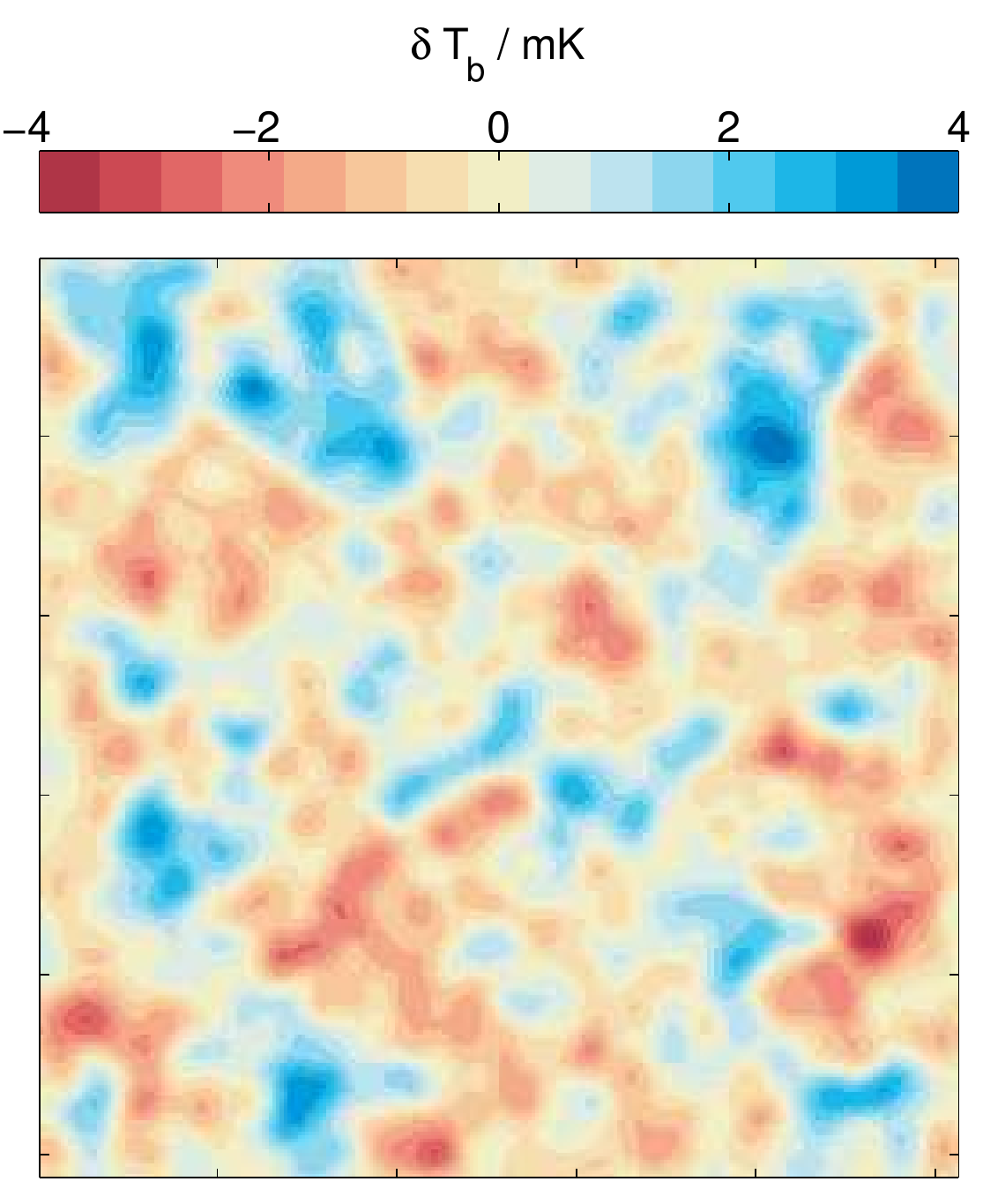} 
\includegraphics[trim=0cm 0cm 0cm 2.9cm,clip=true,width=65mm]{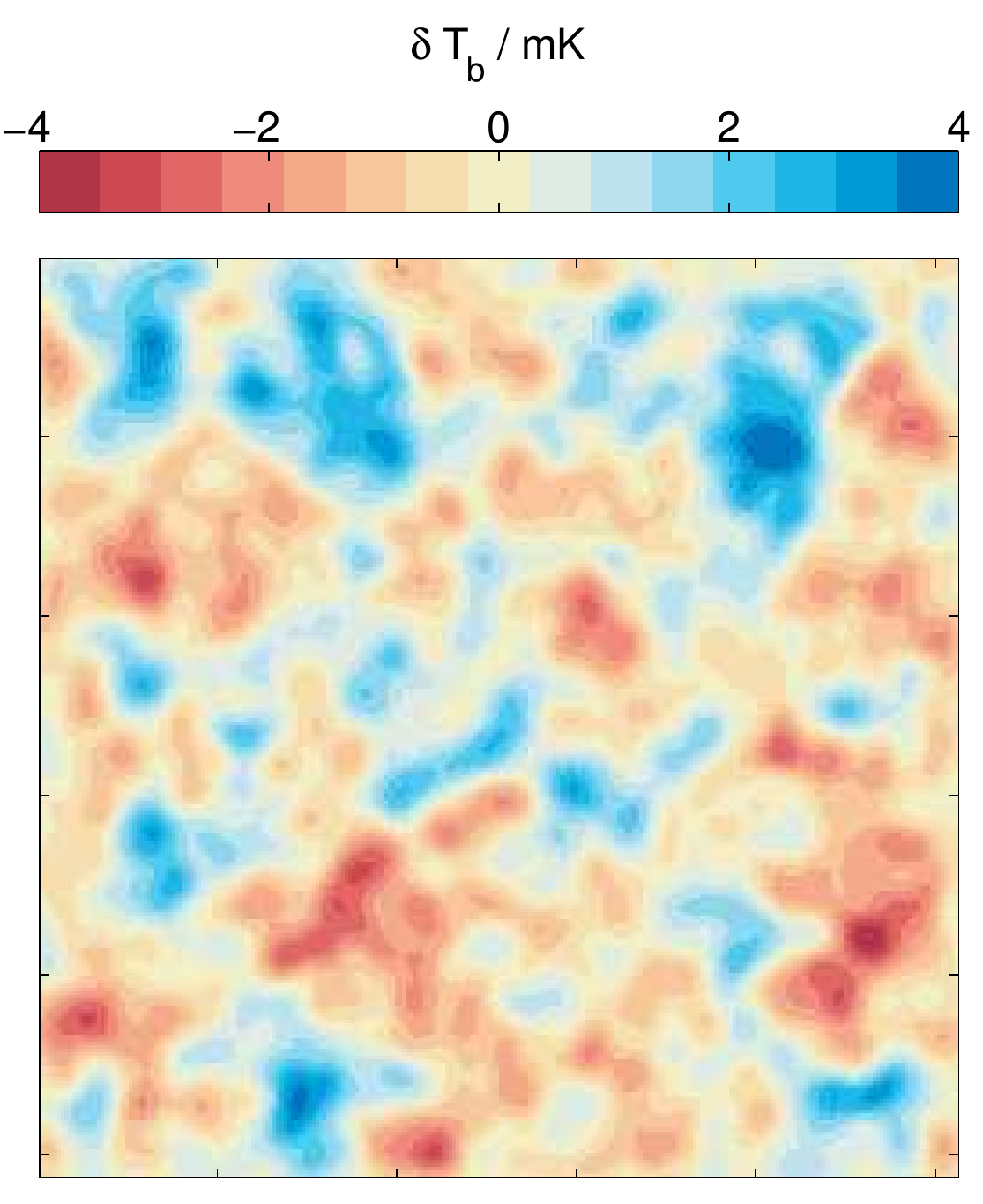} 
\includegraphics[trim=0cm 0cm 0cm 2.9cm,clip=true,width=65mm]{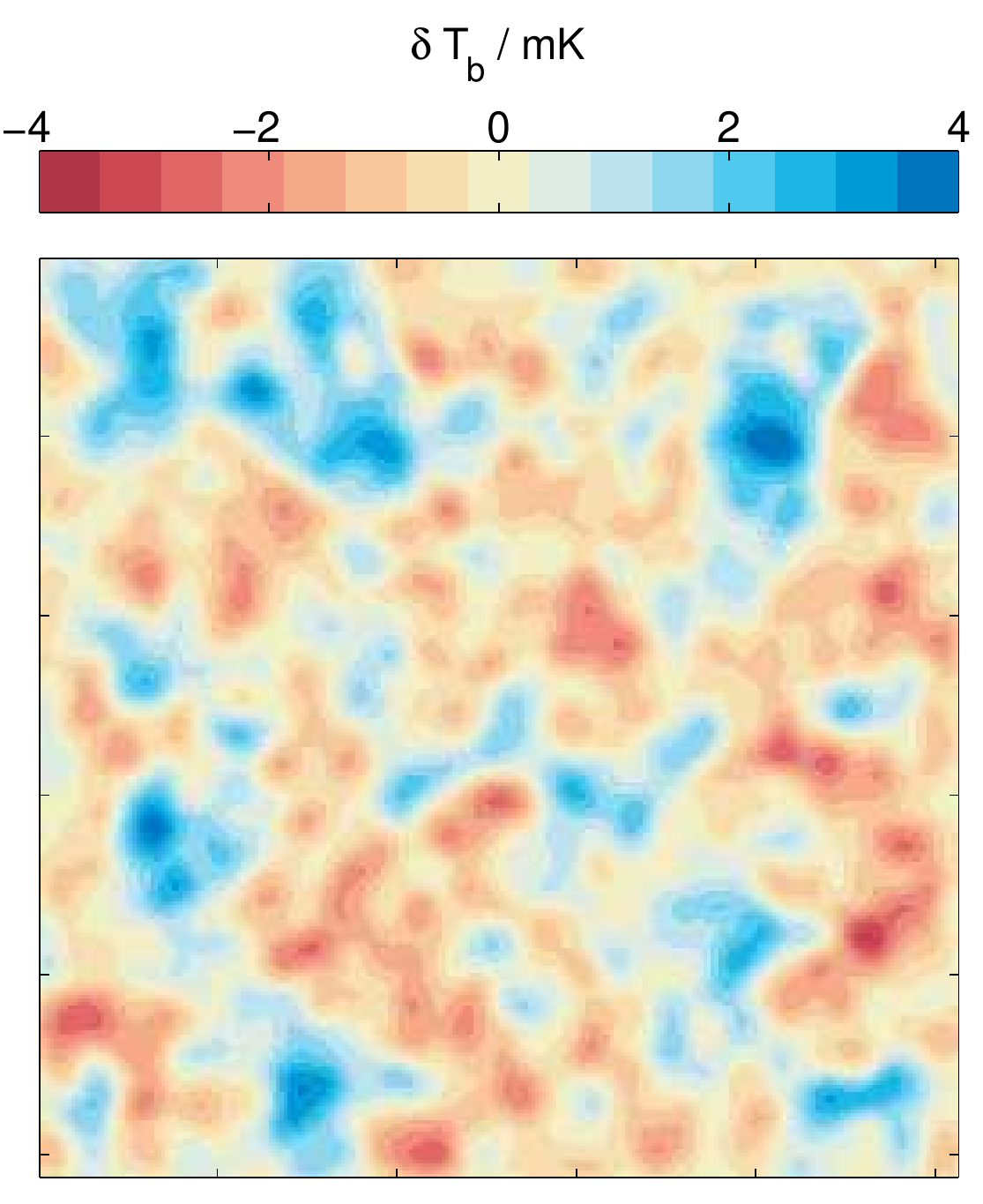} 
\includegraphics[trim=0cm 0cm 0cm 2.9cm,clip=true,width=65mm]{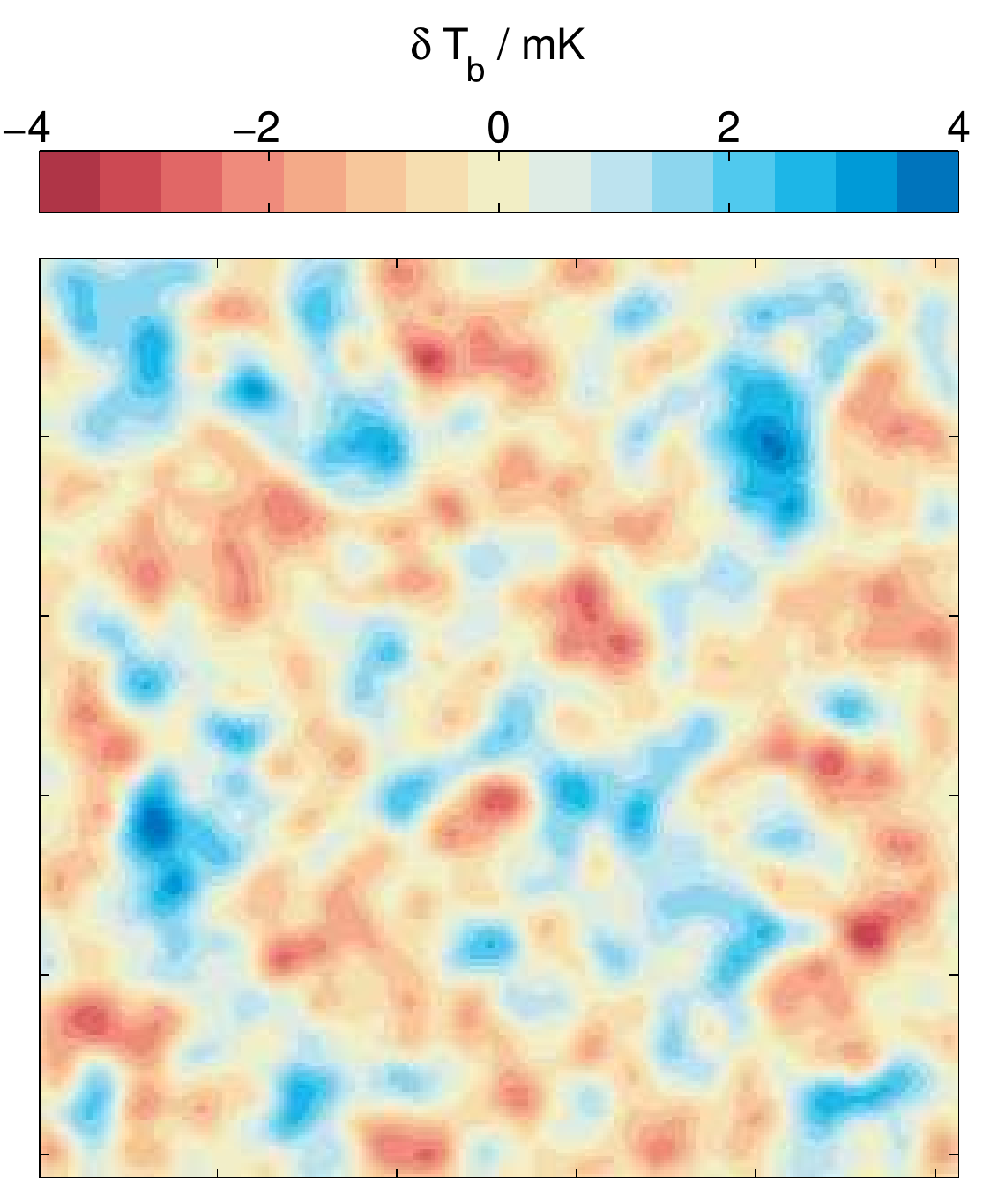}
\includegraphics[width=65mm]{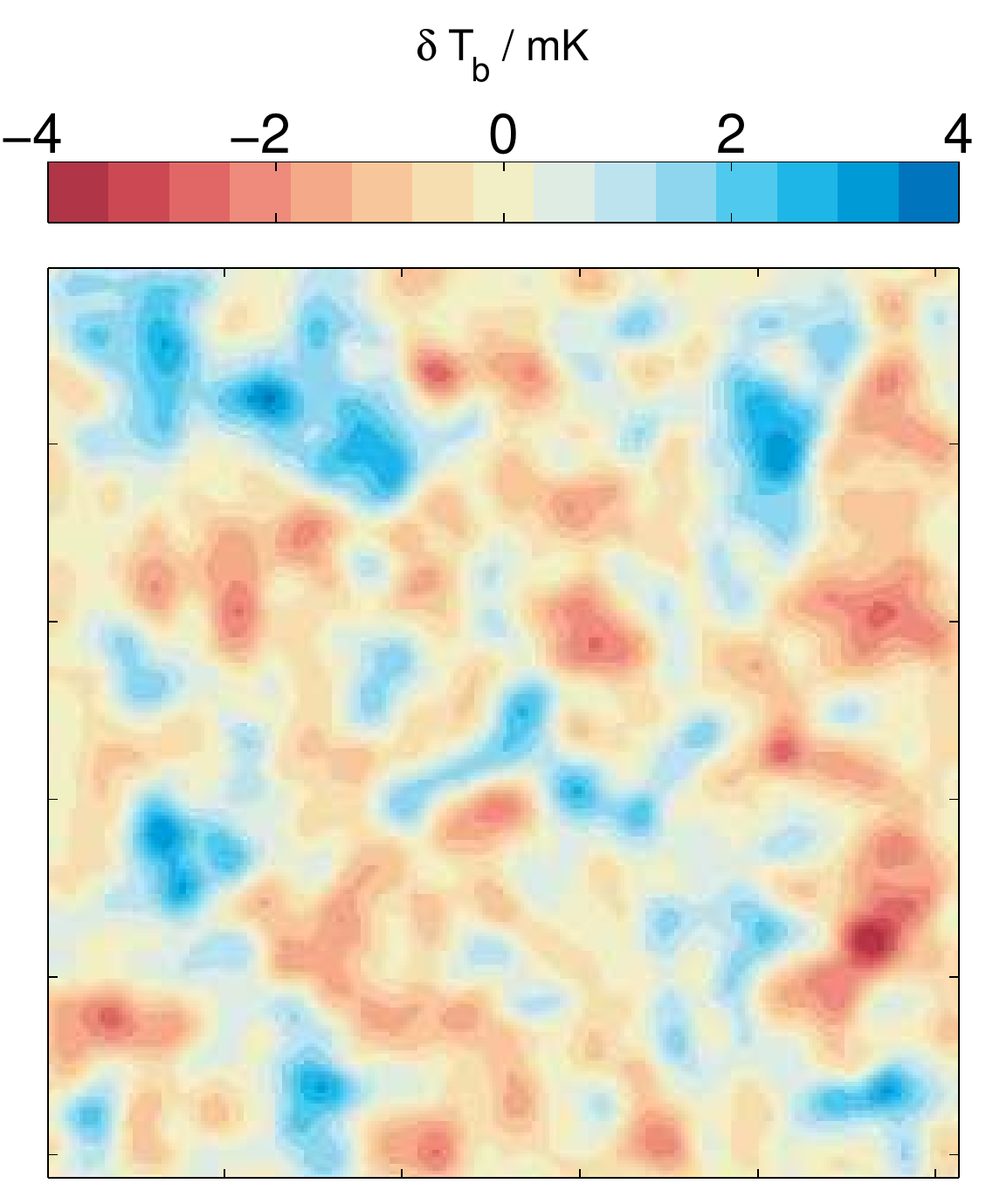}
\includegraphics[width=70mm]{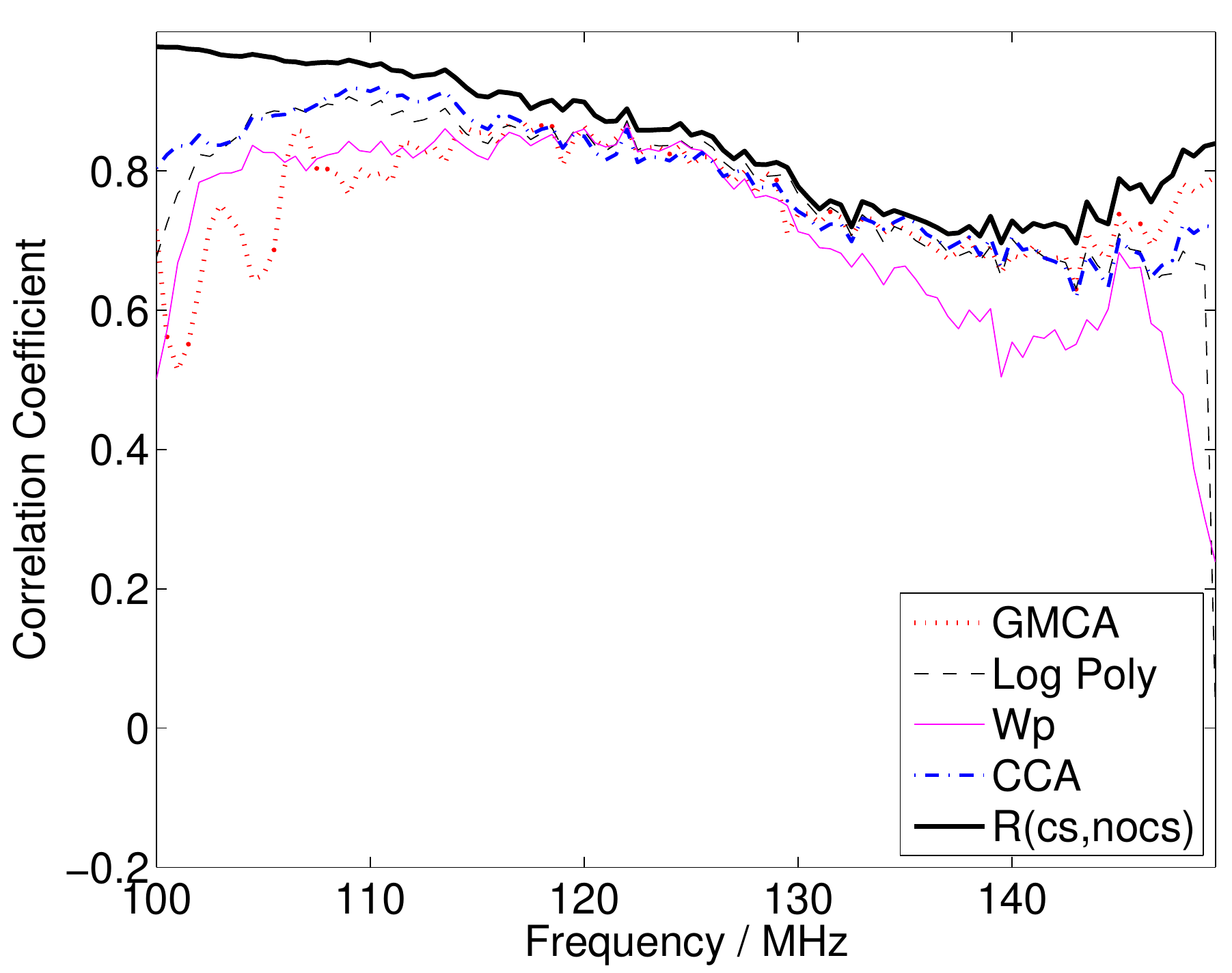} 
\caption{Top two rows, reading order: the smoothed residual maps of S2 at 125\,MHz from the Polynomial, CCA, Wp and GMCA methods. Bottom row, left to right: The smoothed cosmological signal at 125\,MHz and the correlation coefficient relating to residuals vs. cosmological signal.}
\label{fig:im_100}
\end{centering}
\end{figure*}

\begin{figure*}
\begin{centering}
\includegraphics[trim=0cm 0cm 0cm 2.9cm,clip=true,width=65mm]{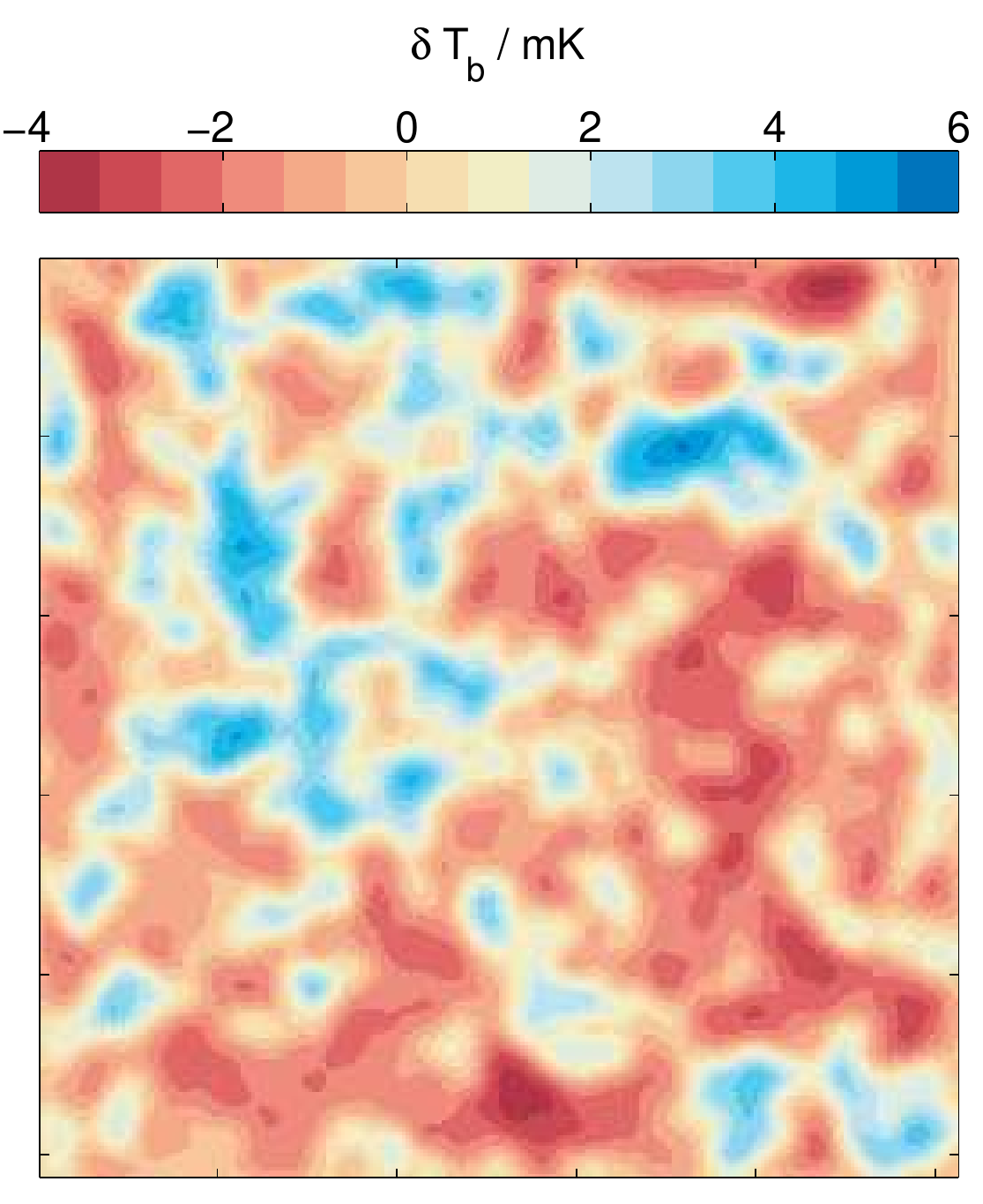} 
\includegraphics[trim=0cm 0cm 0cm 2.9cm,clip=true,width=65mm]{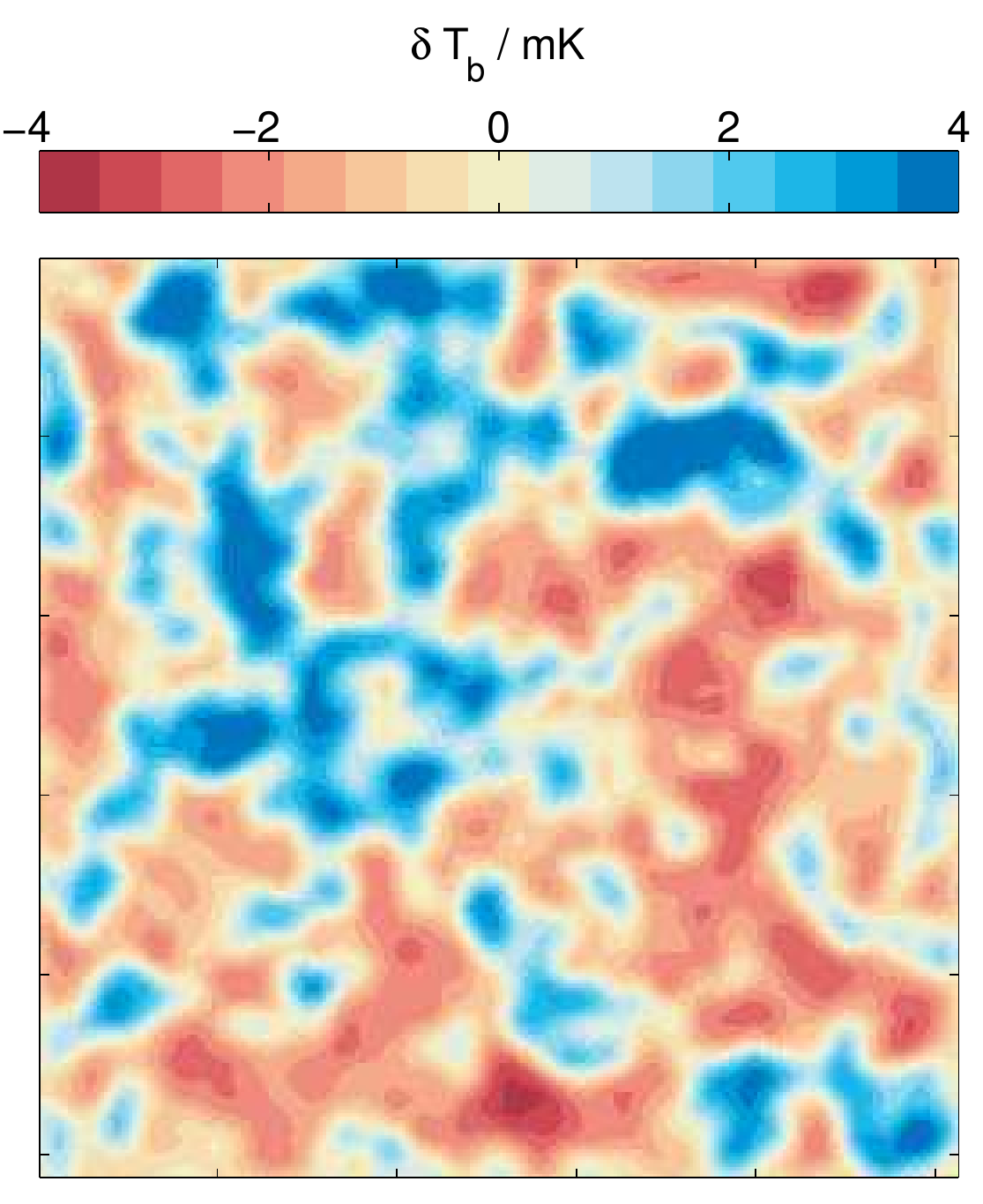} 
\includegraphics[trim=0cm 0cm 0cm 2.9cm,clip=true,width=65mm]{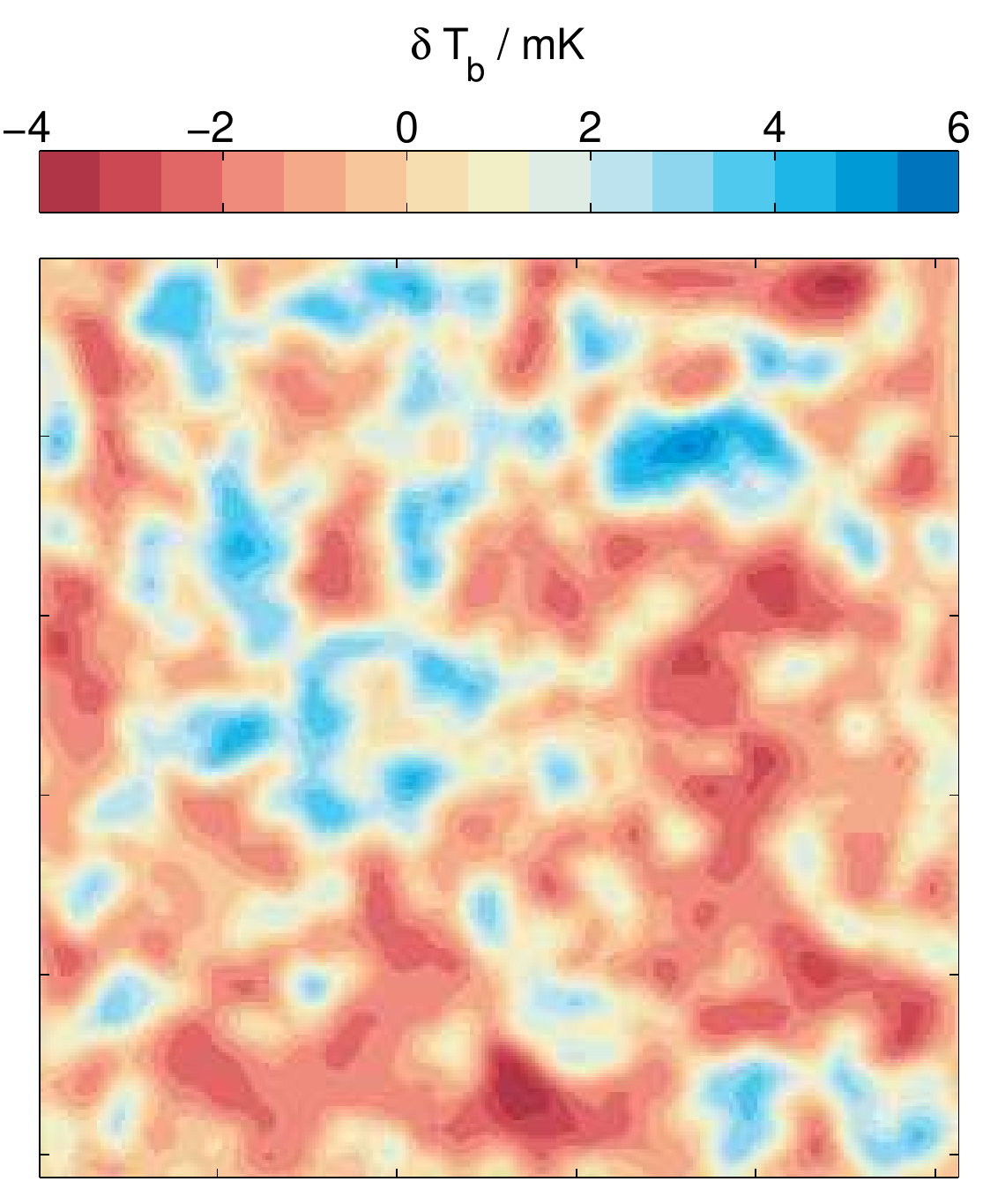} 
\includegraphics[trim=0cm 0cm 0cm 2.9cm,clip=true,width=65mm]{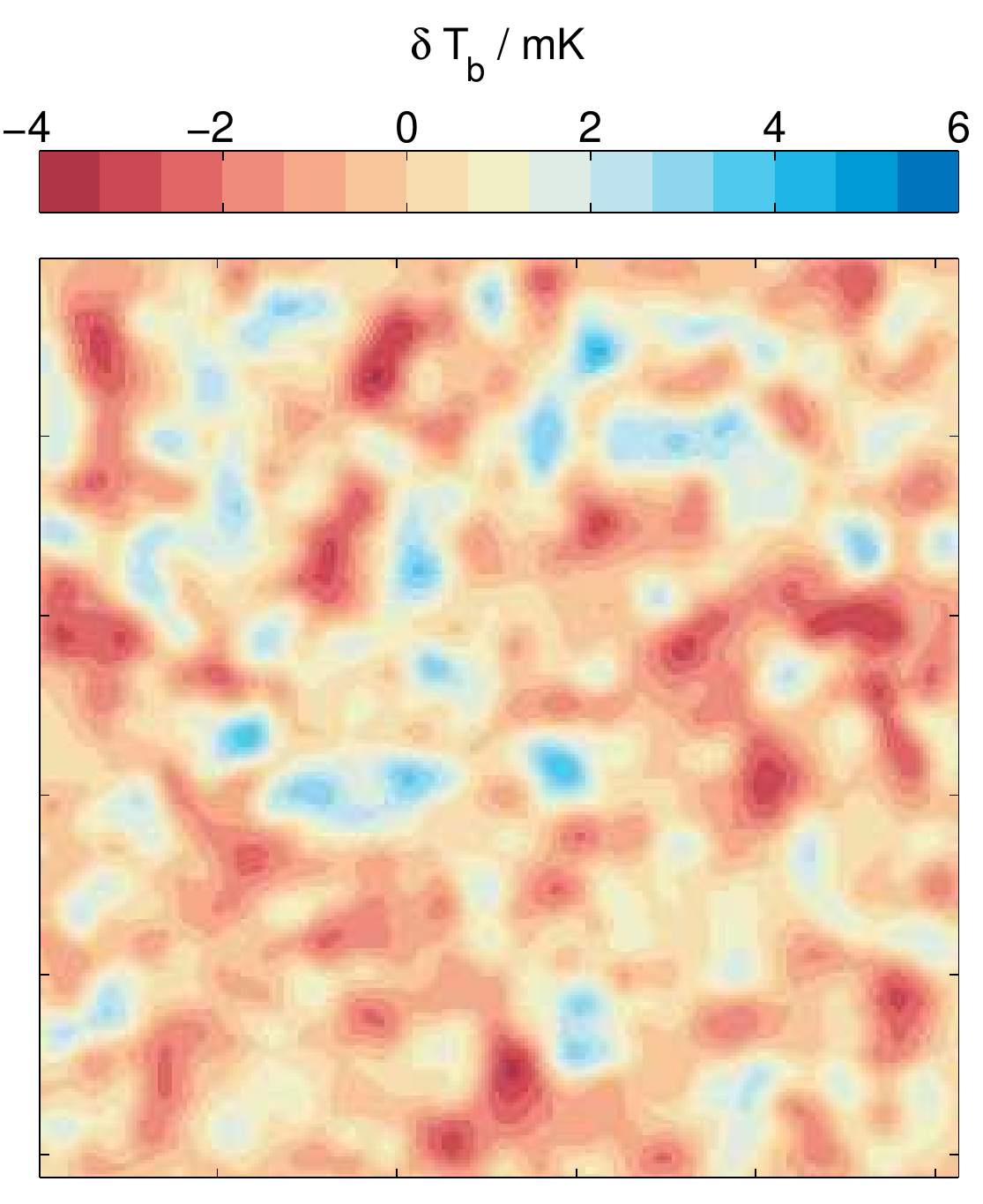}
\includegraphics[width=65mm]{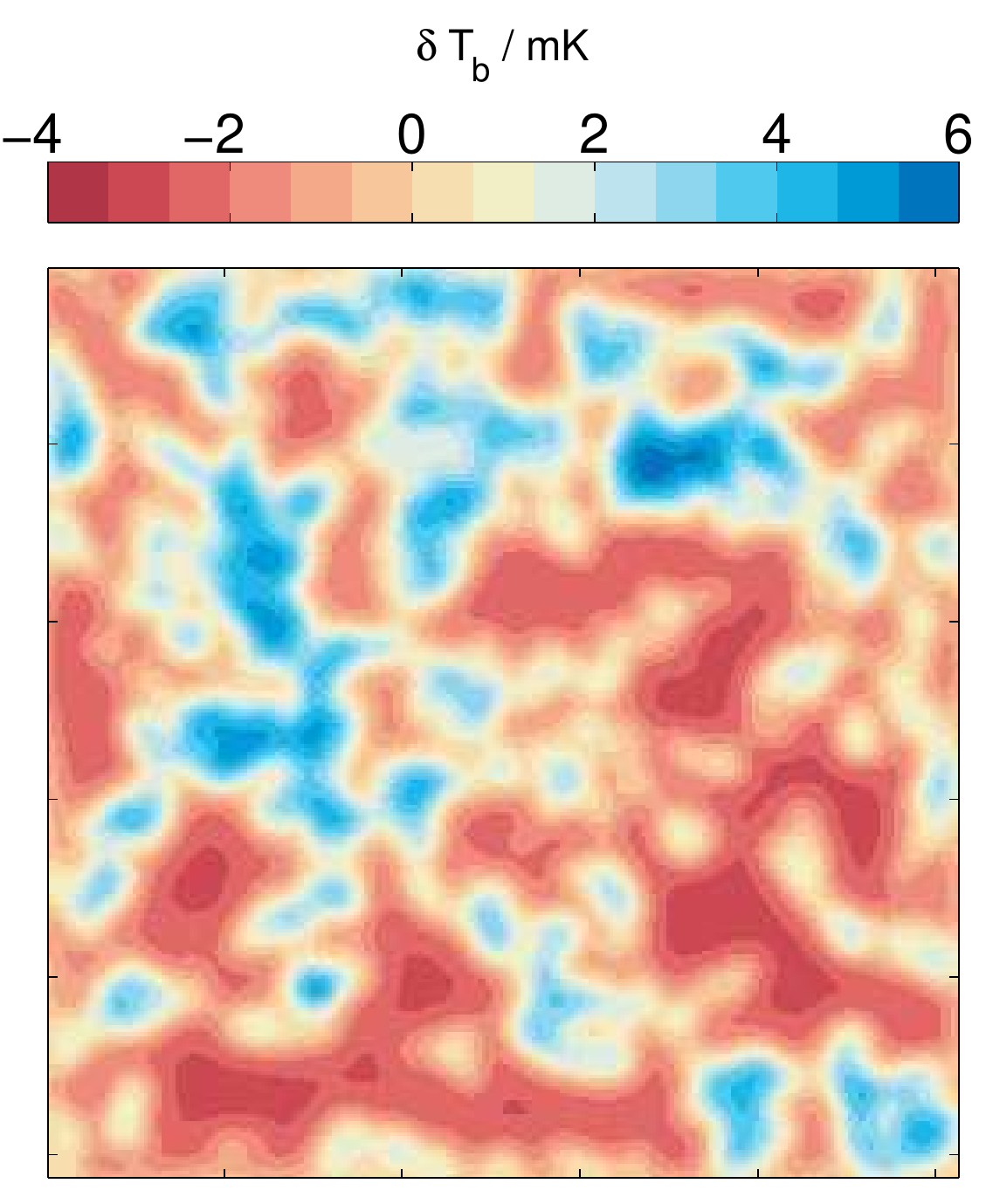}
\includegraphics[width=70mm]{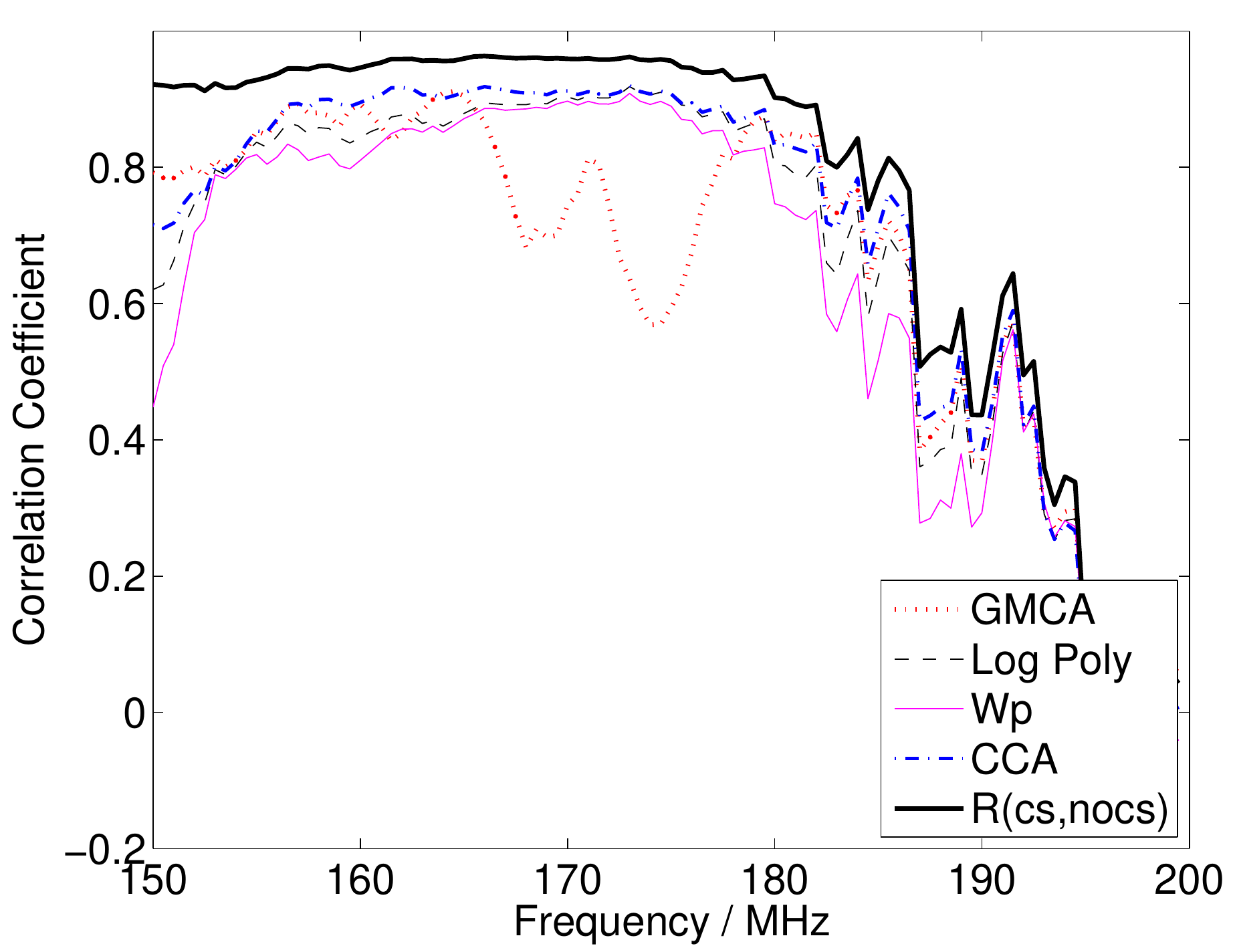} 
\caption{Top two rows, reading order: the smoothed residual maps of S3 at 175\,MHz from the Polynomial, CCA, Wp and GMCA methods. Bottom row, left to right: The smoothed cosmological signal at 175\,MHz and the correlation coefficient relating to residuals vs. cosmological signal.}
\label{fig:im_150}
\end{centering}
\end{figure*}

\subsection{Power Spectrum}
The spherically-averaged 21-cm power spectrum is one of the quantities most readily computed from theory and is a rich source of information on cosmology and on the nature of high-redshift sources \citep[e.g.\ ][and many subsequent works]{BAR05}. It is expected that the SKA will be able measure it with high signal-to-noise across a broad range of scales. Moreover, $k$-space might be considered a natural space in which to study the 21-cm signal, since interferometers natively measure Fourier modes in the plane of the sky, while smooth foregrounds are likely to be more isolated in $k$-space (residing mainly at low $k_{\rm los}$) than in real space. The quality of power spectrum recovery has therefore become a popular metric when comparing foreground removal methods as well as instruments.

Fig.~\ref{fig:3D_ps} shows the spherically averaged power spectrum for a slice in frequency in the centre of S1, S2 and S3. In a given frequency range, this is computed simply by finding the power (magnitude squared of the visibility) in each cell in Fourier space, and then binning this power in spherical shells, i.e. shells with a given $k=\sqrt{k_{\rm perp}^2+k_{\rm los}^2}$. Following convention, we plot $\Delta^2(k)=k^3P(k)/2\pi^2$, which traces the contribution to the variance of a unit interval in $\log(k)$ at $k$.

In each case, the noise power spectrum has been subtracted from the residual power to recover an estimate of the cosmological signal power. Foreground removal mostly yields a reasonable estimate of the cosmological signal, with the possible exception of GMCA in the lowest-redshift slice (in S3). %However, we note that: (1) no attempt has been made to estimate the leakage of the foregrounds into the signal estimate or the loss of signal power into the foreground estimate, as was done by \citet{chapman13}; (2) this bias leads all the methods to underestimate the true power in S3 \citep[see, e.g.\ ][]{petrovic11}; and (3) GMCA is the only method to recover the signal well at all scales in S1 and S2. 
It is not immediately clear why GMCA as opposed to any other method would not perform as well in this frequency segment. The fiducial four-component model of GMCA was chosen rather arbitrarily and ideally all methods should undergo a full Bayesian model selection in order to select the various input parameters (such as the smoothing parameter in Wp smoothing and number of components in the GMCA foreground model). In the S3 panel we have also plotted the power spectrum for GMCA with two and six components in the foreground model. We can see that by changing the number of components in the GMCA foreground model we can achieve a similar fit to the other methods which is perhaps indicative that a more robust method of model selection is needed. It is also worth remembering that the CCA results have undergone an in-built residual minimization which allow it to perform much better than if the minimization were not carried out. This minimization could equally be applied to any method.

We also include in Fig.~\ref{fig:3D_ps} the result of implementing foreground avoidance. By constructing a cylindrical power spectrum in $k_{\rm perp},k_{\rm los}$ we can define a $k_{\rm los}$ bound, below which modes are considered contaminated by foregrounds and above which lies what is termed the `EoR window'. We can then construct a spherical power spectrum as described above but ignoring all modes below this bound. This is the `foreground avoidance' line. We find this bound to be at $k_{\rm los}<10^{-0.93}$, $k_{\rm los}<10^{-0.7}$ and $k_{\rm los}<10^{-0.63} \; \mathrm{Mpc}^{-1}$ for the data at 75, 125 and 175\,MHz respectively (see Fig.~\ref{fig:win}). As one can see, this severely limits the range of scales which can be recovered; however, there are values of $z$ and $k$ for which it performs very well. An optimal power spectrum estimation strategy will likely combine removal and avoidance in some way, with some attempts in this direction having been made by \citet{LIU14b}.

\subsection{Images}
One of the most exciting scientific outcomes of the SKA is the ability to image the EoR and Cosmic Dawn. We now review how the foreground removal methods affect this capability. We present slices from the residual cubes for three different scenarios. In Fig.~\ref{fig:im_50} we take S1 and show the residual slices at 75\,MHz; in Fig.~\ref{fig:im_100} we take S2 and show the residual slices at 125\,MHz and in Fig.~\ref{fig:im_150} we take S3 and show the residual slices at 175\,MHz. Note that, for all images shown, the residual cube has been smoothed with a Gaussian kernel of FWHM eight pixels, which is equivalent to 9.36 arcminutes, in order to mitigate the effect of the instrumental noise. In each figure we also show the correlation coefficient between the smoothed residual cube and the smoothed simulated cosmological signal for the different methods. As we will only have statistical knowledge of the noise, we would not be well motivated in correlating the reconstructed and simulated cosmological signal, and instead also plot an `envelope' in the form of a correlation between the simulated cosmological signal and the simulated cosmological signal combined with the instrumental noise. This provides an upper bound for the best correlation we can expect to see in the data if zero foreground fitting errors were present.  
We see that an impressive image recovery is apparent for all methods, though the extent of that recovery is highly variable with frequency. 

\subsection{Relaxation of foreground smoothness}
We now demonstrate how the methods fare when analysing a cube containing foregrounds which have a random 5\% deviation from the smooth power law along the line of sight. We show the correlation coefficient between the residuals of cube R2 and the cosmological signal in the top-left panel of Fig.~\ref{fig:SKA2}. It is interesting to see the failure of the polynomial method compared to the ability of CCA to recover from a similar failure using the correction method mentioned in Section \ref{subsubsec:CCA}. This correction method could be applied to all approaches and relies on the first order approximation (i.e. that the wiggle is superimposed on a power law) of the foreground being accurate enough. The crucial point to take away is that non-parametric methods, such as GMCA, which do not have a prior on the foreground smoothness, are able to  model the foreground accurately with no extra modelling input by the user. In comparison, the parametric (and indeed non-parametric methods like Wp smoothing which require, as opposed to assume, smoothness of foregrounds) need some form of `extra modelling' such as correction factors or tweaking of the fitting parameters. It is likely that Wp fitting would see a similar improvement to CCA with such correction factors, as the same first-order approximation of smoothness applies.

\subsection{Other SKA configurations}
\begin{figure*}[h]
\begin{centering}
\includegraphics[width=70mm]{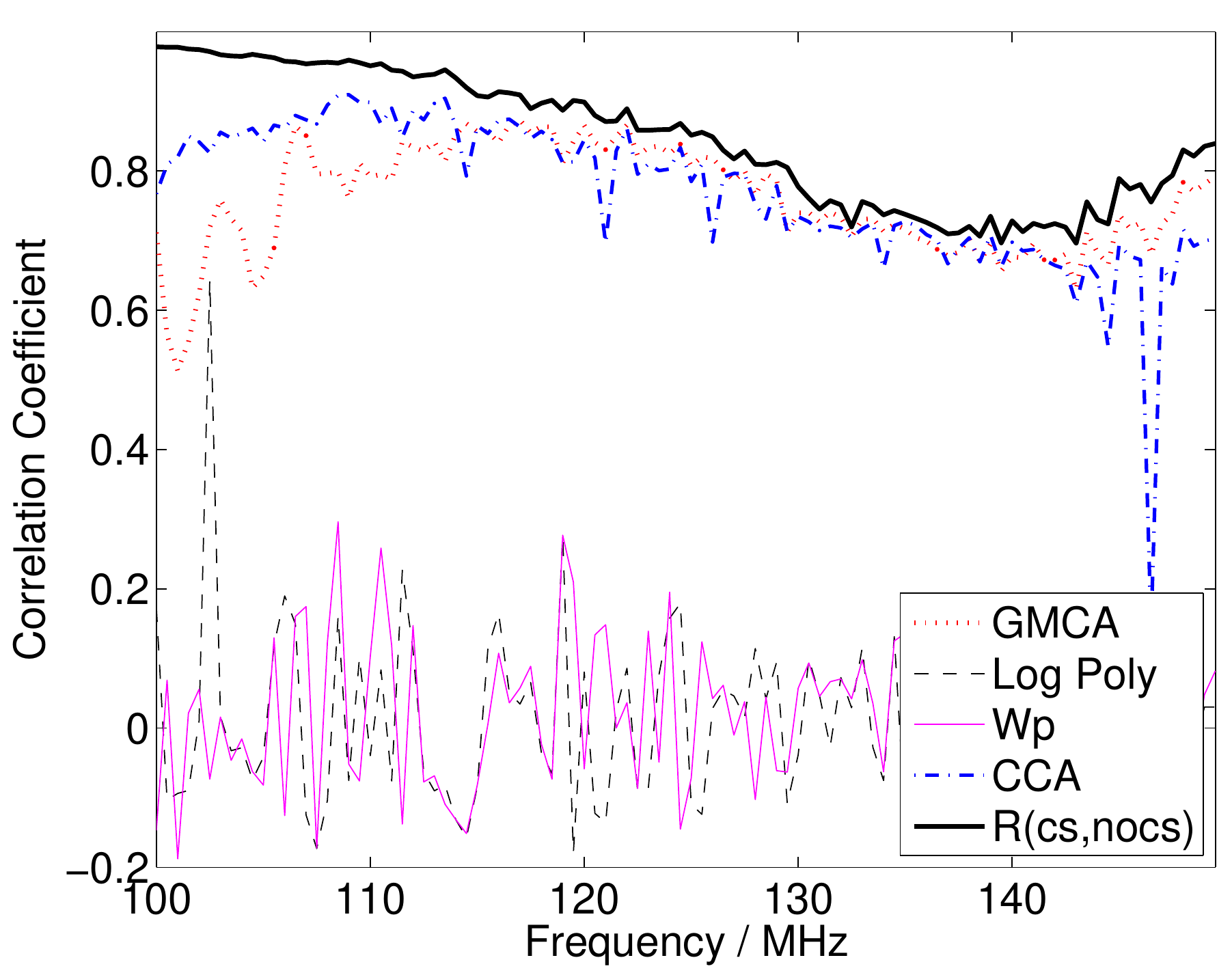} 
\includegraphics[width=70mm]{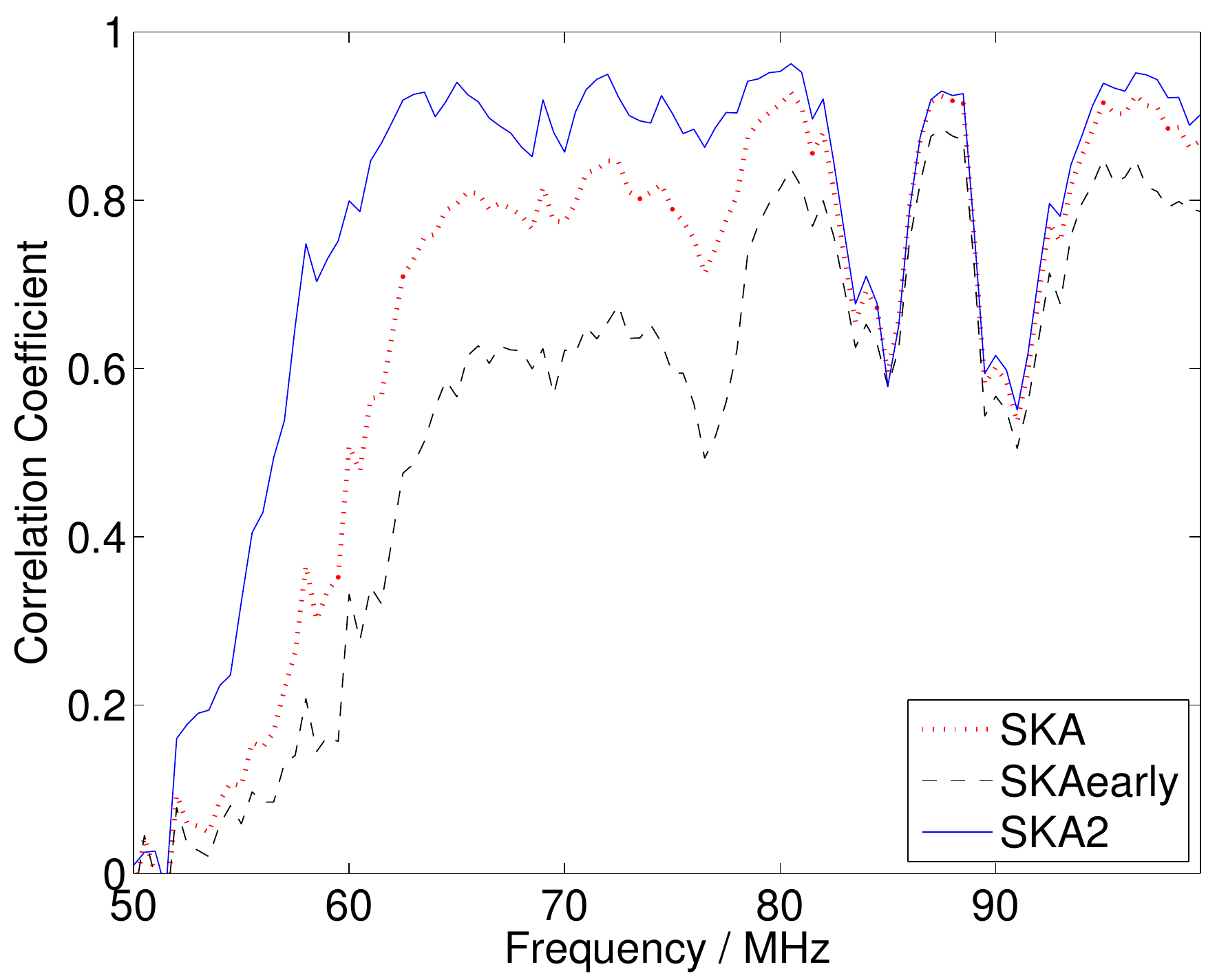} 
\includegraphics[width=70mm]{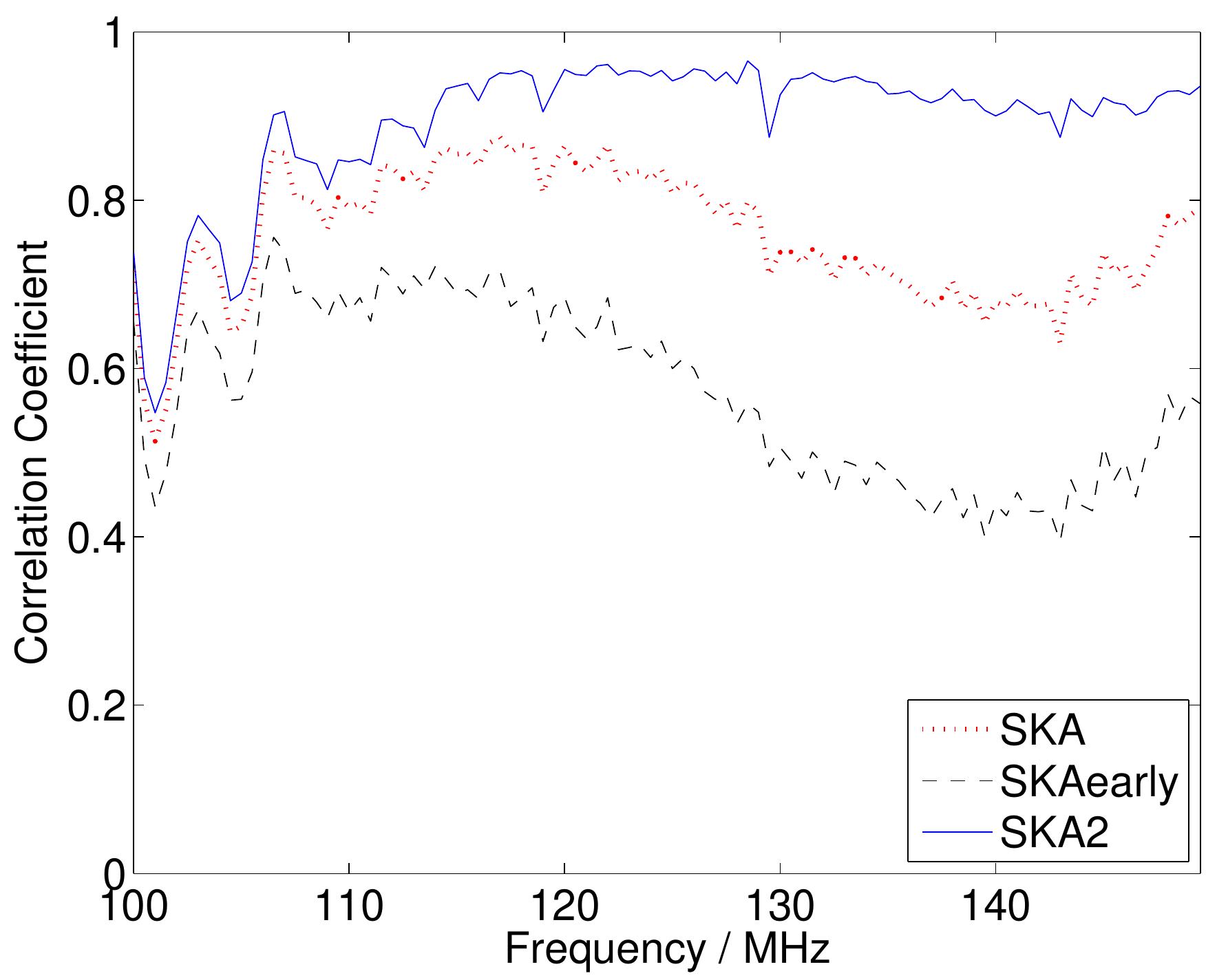} 
\includegraphics[width=70mm]{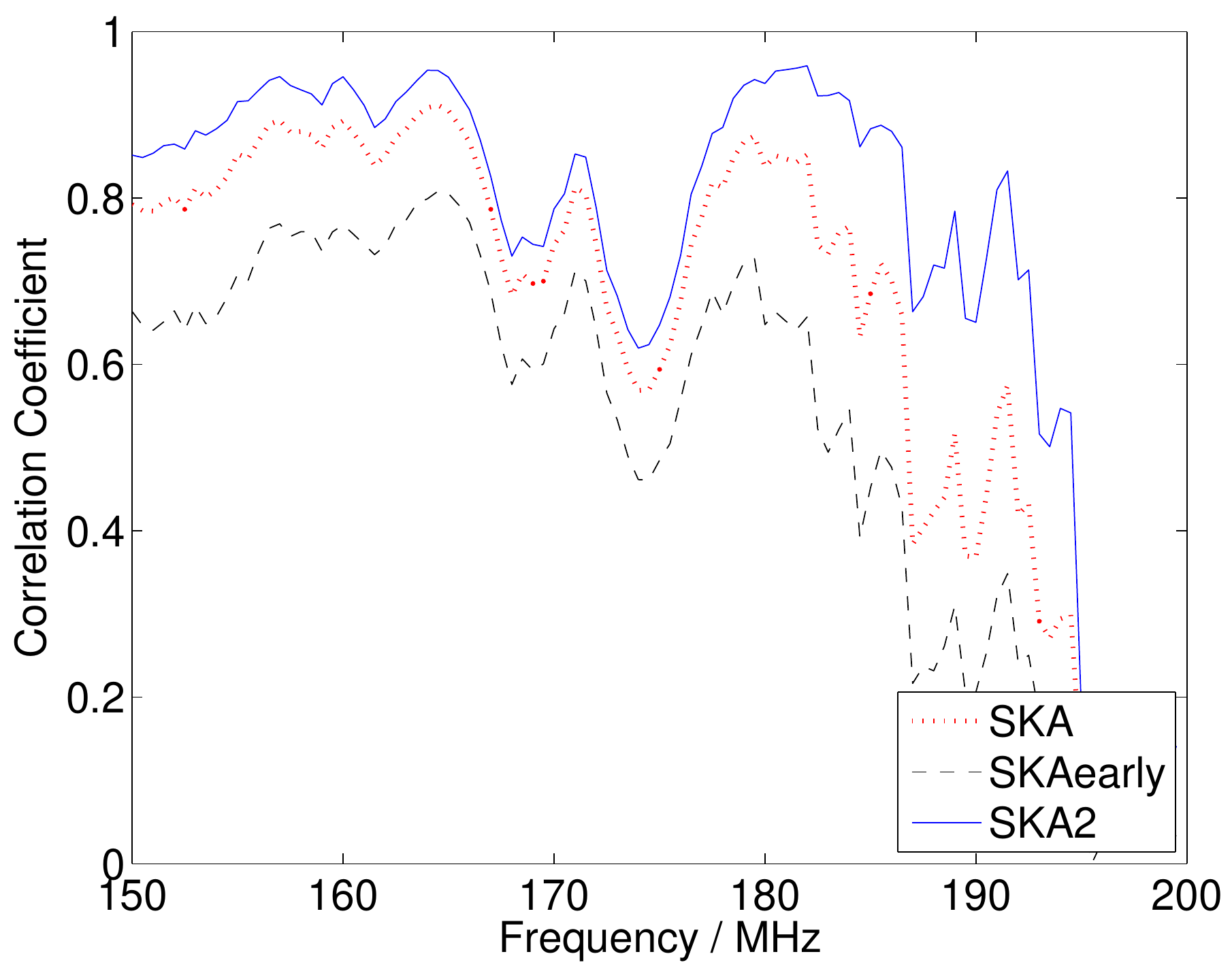} 
\caption{In reading order: The correlation coefficient relating to the R2 cube residuals and the cosmological signal. The correlation coefficient relating to the GMCA residuals for the different SKA scenarios vs. cosmological signal for the S1, S2 and S3 cubes.}
\label{fig:SKA2}
\end{centering}
\end{figure*}

We now look at how one of our results is affected in the case of an early-SKA implementation where the sensitivity is halved and an SKA2 implementation where the sensitivity is quadrupled. We do this for only one method, GMCA, for clarity and conciseness. We show the correlation coefficient between the recovered maps in the three remaining panels of Fig.~\ref{fig:SKA2}. 

As the instrumental noise level decreases, the recovery is greatly improved. The general shape of the curves remains similar, suggesting that there is significant contribution by foreground fitting errors which affect the correlation from slice to slice. There are indeed slices in the S1 cube, for example at 85\,MHz, which produce the same correlation independent of noise level, suggesting foreground leakage is the dominant cause of error at those frequencies.  

\section{Conclusions}
\label{sec:conc}
\begin{itemize}
\item We have applied a suite of foreground removal methods to a state-of-the-art simulation of SKA Phase 1 observations and analysed the recovered cosmological signal by using different statistics.
\item The variance of the 21-cm signal is well recovered over a broad range of frequencies by all methods. While setting the entire bandwidth to a common resolution results in the loss of a lot of spatial information at high frequency, there is no obvious advantage in the accuracy of the recovered variance. We therefore opt for a compromise of setting several segments to common resolution.
\item Foreground removal methods generally yield a reasonable estimate of the spherical power spectrum of the cosmological signal. Foreground avoidance severely limits the range of scales for which the power spectrum can be recovered, however in such a limited range the results are good.
\item We obtain an impressive recovery of the images for all methods, the quality of which however varies with frequency. 
\item The relaxation of the hypothesis of foreground smoothness does not affect the performance of GMCA, which is non-parametric, while it affects polynomial fitting and Wp smoothing. However, as shown by the CCA, the quality of the results can be restored with some extra modelling, at least as long as the smooth model adopted is a reasonable approximation of the foreground spectrum.
\item As the instrumental noise level decreases, the recovery of the signal is greatly improved. For some frequencies (85\,MHz for example) the results are much more similar, which suggests that foreground subtraction errors in these cases are dominant. 

\end{itemize}

\section{Acknowledgments}
EC would like to thank Jonathan Pritchard and the SKA-EoR working group for useful discussions. GH acknowledges funding from the People Programme (Marie Curie Actions) of the European Union's Seventh Framework Programme (FP7/2007-2013) under REA Grant Agreement No 327999. AB acknowledges support from the European Research Council under the EC FP7 grant number 280127.
\newpage
\bibliographystyle{apj}
\bibliography{CDEOR_FGremoval_2.0}

\begin{thebibliography}{53}
\expandafter\ifx\csname natexlab\endcsname\relax\def\natexlab#1{#1}\fi

\bibitem[{{Barkana} \& {Loeb}(2005)}]{BAR05}
{Barkana}, R. \& {Loeb}, A. 2005, ApJL, 624, L65

\bibitem[{Bernardi {et~al.}(2009)Bernardi, de~Bruyn, Brentjens, Ciardi, Harker,
  Jeli{\'c}, Koopmans, Labropoulos, Offringa, Pandey, Schaye, Thomas,
  Yatawatta, \& Zaroubi}]{bernardi09}
Bernardi, G., de~Bruyn, A.~G., Brentjens, M.~A., Ciardi, B., Harker, G.,
  Jeli{\'c}, V., Koopmans, L. V.~E., Labropoulos, P., Offringa, A., Pandey,
  V.~N., Schaye, J., Thomas, R.~M., Yatawatta, S., \& Zaroubi, S. 2009, A\&A,
  500, 965

\bibitem[{{Bernardi} {et~al.}(2010){Bernardi}, {de Bruyn}, {Harker},
  {Brentjens}, {Ciardi}, {Jeli{\'c}}, {Koopmans}, {Labropoulos}, {Offringa},
  {Pandey}, {Schaye}, {Thomas}, {Yatawatta}, \& {Zaroubi}}]{bernardi2010}
{Bernardi}, G., {de Bruyn}, A.~G., {Harker}, G., {Brentjens}, M.~A., {Ciardi},
  B., {Jeli{\'c}}, V., {Koopmans}, L.~V.~E., {Labropoulos}, P., {Offringa}, A.,
  {Pandey}, V.~N., {Schaye}, J., {Thomas}, R.~M., {Yatawatta}, S., \&
  {Zaroubi}, S. 2010, A\&A, 522, A67

\bibitem[{Bernardi {et~al.}(2013)Bernardi, Greenhill, Mitchell, Ord, Hazelton,
  Gaensler, de~Oliveira-Costa, Morales, Shankar, Subrahmanyan, Wayth, Lenc,
  Williams, Arcus, Arora, Barnes, Bowman, Briggs, Bunton, Cappallo, Corey,
  Deshpande, deSouza, Emrich, Goeke, Herne, Hewitt, Johnston-Hollitt, Kaplan,
  Kasper, Kincaid, Koenig, Kratzenberg, Lonsdale, Lynch, McWhirter, Morgan,
  Oberoi, Pathikulangara, Prabu, Remillard, Rogers, Roshi, Salah, Sault,
  Srivani, Stevens, Tingay, Waterson, Webster, Whitney, Williams, \&
  Wyithe}]{bernardi13}
Bernardi, G., Greenhill, L.~J., Mitchell, D.~A., Ord, S.~M., Hazelton, B.~J.,
  Gaensler, B.~M., de~Oliveira-Costa, A., Morales, M.~F., Shankar, N.~U.,
  Subrahmanyan, R., Wayth, R.~B., Lenc, E., Williams, C.~L., Arcus, W., Arora,
  B.~S., Barnes, D.~G., Bowman, J.~D., Briggs, F.~H., Bunton, J.~D., Cappallo,
  R.~J., Corey, B.~E., Deshpande, A., deSouza, L., Emrich, D., Goeke, R.,
  Herne, D., Hewitt, J.~N., Johnston-Hollitt, M., Kaplan, D., Kasper, J.~C.,
  Kincaid, B.~B., Koenig, R., Kratzenberg, E., Lonsdale, C.~J., Lynch, M.~J.,
  McWhirter, S.~R., Morgan, E., Oberoi, D., Pathikulangara, J., Prabu, T.,
  Remillard, R.~A., Rogers, A. E.~E., Roshi, A., Salah, J.~E., Sault, R.~J.,
  Srivani, K.~S., Stevens, J., Tingay, S.~J., Waterson, M., Webster, R.~L.,
  Whitney, A.~R., Williams, A., \& Wyithe, J. S.~B. 2013, The Astrophysical
  Journal, 771, 105

\bibitem[{Bobin {et~al.}(2008{\natexlab{a}})Bobin, Moudden, Starck, Fadili, \&
  Aghanim}]{bobin08a}
Bobin, J., Moudden, Y., Starck, J.-L., Fadili, J., \& Aghanim, N.
  2008{\natexlab{a}}, Statistical Methodology, 5, 307

\bibitem[{Bobin {et~al.}(2007)Bobin, Starck, Fadili, \& Moudden}]{bobin07}
Bobin, J., Starck, J.-L., Fadili, J., \& Moudden, Y. 2007, IEEE Transactions on
  Image Processing, 16, 2662

\bibitem[{Bobin {et~al.}(2008{\natexlab{b}})Bobin, Starck, Moudden, \&
  Fadili}]{bobin08b}
Bobin, J., Starck, J.-L., Moudden, Y., \& Fadili, M.~J. 2008{\natexlab{b}}, in
  (Elsevier), 221 -- 302

\bibitem[{Bobin {et~al.}(2013)Bobin, {Starck, J.-L.}, {Sureau, F.}, \& {Basak,
  S.}}]{bobin12}
Bobin, J., {Starck, J.-L.}, {Sureau, F.}, \& {Basak, S.} 2013, A\&A, 550, A73

\bibitem[{{Bobin} {et~al.}(2014){Bobin}, {Sureau}, {Starck}, {Rassat}, \&
  {Paykari}}]{GMCA_PR1}
{Bobin}, J., {Sureau}, F., {Starck}, J.-L., {Rassat}, A., \& {Paykari}, P.
  2014, A\&A, 563

\bibitem[{Bonaldi \& Brown(2015)}]{bonaldibrown}
Bonaldi, A. \& Brown, M.~L. 2015, Monthly Notices of the Royal Astronomical
  Society, 447, 1973

\bibitem[{{Bonaldi} \& {Ricciardi}(2012)}]{special}
{Bonaldi}, A. \& {Ricciardi}, S. 2012, Advances in Astronomy

\bibitem[{{Bonaldi} {et~al.}(2007){Bonaldi}, {Ricciardi}, {Leach}, {Stivoli},
  {Baccigalupi}, \& {de Zotti}}]{bonaldi2007}
{Bonaldi}, A., {Ricciardi}, S., {Leach}, S., {Stivoli}, F., {Baccigalupi}, C.,
  \& {de Zotti}, G. 2007, MNRAS, 382, 1791

\bibitem[{Bowman {et~al.}(2006)Bowman, Morales, \& Hewitt}]{bowman09}
Bowman, J.~D., Morales, M.~F., \& Hewitt, J.~N. 2006, ApJ, 638, 20

\bibitem[{Chapman {et~al.}(2012)Chapman, Abdalla, Harker, Jelic, Labropoulos,
  Zaroubi, Brentjens, de~Bruyn, \& Koopmans}]{chapman12}
Chapman, E., Abdalla, F., Harker, G., Jelic, V., Labropoulos, P., Zaroubi, S.,
  Brentjens, M.~A., de~Bruyn, A.~G., \& Koopmans, L. V.~E. 2012, MNRAS, 423,
  2518

\bibitem[{{Chapman} {et~al.}(2013){Chapman}, {Abdalla}, {Bobin}, {Starck},
  {Harker}, {Jeli{\'c}}, {Labropoulos}, {Zaroubi}, {Brentjens}, {de Bruyn}, \&
  {Koopmans}}]{chapman13}
{Chapman}, E., {Abdalla}, F.~B., {Bobin}, J., {Starck}, J.-L., {Harker}, G.,
  {Jeli{\'c}}, V., {Labropoulos}, P., {Zaroubi}, S., {Brentjens}, M.~A., {de
  Bruyn}, A.~G., \& {Koopmans}, L.~V.~E. 2013, MNRAS, 429, 165

\bibitem[{Chapman {et~al.}(2014)Chapman, Zaroubi, \& Abdalla}]{chapman14}
Chapman, E., Zaroubi, S., \& Abdalla, F. 2014, preprint (astro-ph/1408.4695)

\bibitem[{{Datta} {et~al.}(2010){Datta}, {Bowman}, \& {Carilli}}]{datta10}
{Datta}, A., {Bowman}, J.~D., \& {Carilli}, C.~L. 2010, ApJ, 724, 526

\bibitem[{{Di Matteo} {et~al.}(2002){Di Matteo}, {Perna}, {Abel}, \&
  {Rees}}]{dimatteo02}
{Di Matteo}, T., {Perna}, R., {Abel}, T., \& {Rees}, M. 2002, ApJ, 564, 576

\bibitem[{{{Di Matteo}, T. and {Ciardi}, B. and {Miniati},
  F.}(2004)}]{dimatteo04}
{{Di Matteo}, T. and {Ciardi}, B. and {Miniati}, F.} 2004, MNRAS, 355, 1053

\bibitem[{Dillon {et~al.}(2014)Dillon, Liu, Williams, Hewitt, Tegmark, Morgan,
  Levine, Morales, Tingay, Bernardi, Bowman, Briggs, Cappallo, Emrich,
  Mitchell, Oberoi, Prabu, Wayth, \& Webster}]{dillon13}
Dillon, J.~S., Liu, A., Williams, C.~L., Hewitt, J.~N., Tegmark, M., Morgan,
  E.~H., Levine, A.~M., Morales, M.~F., Tingay, S.~J., Bernardi, G., Bowman,
  J., Briggs, F.~H., Cappallo, R.~C., Emrich, D., Mitchell, D.~A., Oberoi, D.,
  Prabu, T., Wayth, R., \& Webster, R.~L. 2014, Phys. Rev. D, 89, 023002

\bibitem[{{Gleser} {et~al.}(2008){Gleser}, {Nusser}, \& {Benson}}]{gleser08}
{Gleser}, L., {Nusser}, A., \& {Benson}, A.~J. 2008, MNRAS, 391, 383

\bibitem[{Gnedin \& Shaver(2004)}]{gnedin04}
Gnedin, N.~Y. \& Shaver, P.~A. 2004, ApJ, 608, 611

\bibitem[{{Gu} {et~al.}(2013){Gu}, {Xu}, {Wang}, {An}, \& {Chen}}]{GU13}
{Gu}, J., {Xu}, H., {Wang}, J., {An}, T., \& {Chen}, W. 2013, ApJ, 773, 38

\bibitem[{{Harker} {et~al.}(2009){Harker}, {Zaroubi}, {Bernardi}, {Brentjens},
  {de Bruyn}, {Ciardi}, {Jeli{\'c}}, {Koopmans}, {Labropoulos}, {Mellema},
  {Offringa}, {Pandey}, {Schaye}, {Thomas}, \& {Yatawatta}}]{NONPAR_09}
{Harker}, G., {Zaroubi}, S., {Bernardi}, G., {Brentjens}, M.~A., {de Bruyn},
  A.~G., {Ciardi}, B., {Jeli{\'c}}, V., {Koopmans}, L.~V.~E., {Labropoulos},
  P., {Mellema}, G., {Offringa}, A., {Pandey}, V.~N., {Schaye}, J., {Thomas},
  R.~M., \& {Yatawatta}, S. 2009, MNRAS, 397, 1138

\bibitem[{Hyv{\"a}rinen(1999)}]{hyvarinen99}
Hyv{\"a}rinen, A. 1999, Neural Networks, IEEE Transactions on, 10, 626

\bibitem[{Hyv{\"a}rinen {et~al.}(2001)Hyv{\"a}rinen, Karhunen, \&
  Oja}]{hyvarinen01}
Hyv{\"a}rinen, A., Karhunen, J., \& Oja, E. 2001, {\em Independent Component
  Analysis} (John Wiley and Sons)

\bibitem[{Jeli\'{c} {et~al.}(2010)Jeli\'{c}, Zaroubi, Labropoulos, Bernardi,
  de~Bruyn, \& Koopmans}]{jelic10}
Jeli\'{c}, V., Zaroubi, S., Labropoulos, P., Bernardi, G., de~Bruyn, A.~G., \&
  Koopmans, L. V.~E. 2010, MNRAS, 409, 1647

\bibitem[{{Jeli{\'c}} {et~al.}(2008){Jeli{\'c}}, {Zaroubi}, {Labropoulos},
  {Thomas}, {Bernardi}, {Brentjens}, {de Bruyn}, {Ciardi}, {Harker},
  {Koopmans}, {Pandey}, {Schaye}, \& {Yatawatta}}]{jelic2008}
{Jeli{\'c}}, V., {Zaroubi}, S., {Labropoulos}, P., {Thomas}, R.~M., {Bernardi},
  G., {Brentjens}, M.~A., {de Bruyn}, A.~G., {Ciardi}, B., {Harker}, G.,
  {Koopmans}, L.~V.~E., {Pandey}, V.~N., {Schaye}, J., \& {Yatawatta}, S. 2008,
  MNRAS, 389, 1319

\bibitem[{Liu {et~al.}(2014a)Liu, Parsons, \& Trott}]{liu14a}
Liu, A., Parsons, A.~R., \& Trott, C.~M. 2014a, Phys. Rev. D, 90, 023018

\bibitem[{{Liu} {et~al.}(2014b){Liu}, {Parsons}, \& {Trott}}]{LIU14b}
{Liu}, A., {Parsons}, A.~R., \& {Trott}, C.~M. 2014b, Phys. Rev. D, 90, 023019

\bibitem[{Liu \& Tegmark(2011)}]{liu11}
Liu, A. \& Tegmark, M. 2011, Phys. Rev. D, 83, 103006

\bibitem[{{Liu} {et~al.}(2009){Liu}, {Tegmark}, {Bowman}, {Hewitt}, \&
  {Zaldarriaga}}]{liu09}
{Liu}, A., {Tegmark}, M., {Bowman}, J., {Hewitt}, J., \& {Zaldarriaga}, M.
  2009, MNRAS, 398, 401

\bibitem[{{M{\"a}chler}(1989)}]{MAC89}
{M{\"a}chler}, M. 1989, PhD thesis, ETH Z\"urich

\bibitem[{{M{\"a}chler}(1993)}]{MAC93}
---. 1993, {\em Very smooth nonparametric curve estimation by penalizing change
  of curvature}, {Research report}~71, {Seminar f\"ur Statistik ETH Z\"urich}

\bibitem[{{M{\"a}chler}(1995)}]{MAC95}
---. 1995, {Annals of Statistics}, 23, 1496

\bibitem[{{McQuinn} {et~al.}(2006){McQuinn}, {Zahn}, {Zaldarriaga},
  {Hernquist}, \& {Furlanetto}}]{mcquinn06}
{McQuinn}, M., {Zahn}, O., {Zaldarriaga}, M., {Hernquist}, L., \& {Furlanetto},
  S.~R. 2006, ApJ, 653, 815

\bibitem[{{Morales} {et~al.}(2006){Morales}, {Bowman}, \& {Hewitt}}]{morales06}
{Morales}, M.~F., {Bowman}, J.~D., \& {Hewitt}, J.~N. 2006, ApJ, 648, 767

\bibitem[{Morales {et~al.}(2012)Morales, Hazelton, Sullivan, \&
  Beardsley}]{morales12}
Morales, M.~F., Hazelton, B., Sullivan, I., \& Beardsley, A. 2012, The
  Astrophysical Journal, 752, 137

\bibitem[{Oh \& Mack(2003)}]{oh03}
Oh, S.~P. \& Mack, K.~J. 2003, MNRAS, 346, 871

\bibitem[{Parsons {et~al.}(2012)Parsons, Pober, Aguirre, Carilli, Jacobs, \&
  Moore}]{parsons12}
Parsons, A.~R., Pober, J.~C., Aguirre, J.~E., Carilli, C.~L., Jacobs, D.~C., \&
  Moore, D.~F. 2012, The Astrophysical Journal, 756, 165

\bibitem[{{Patil} {et~al.}(2014){Patil}, {Zaroubi}, {Chapman}, {Jeli{\'c}},
  {Harker}, {Abdalla}, {Asad}, {Bernardi}, {Brentjens}, {de Bruyn}, {Bus},
  {Ciardi}, {Daiboo}, {Fernandez}, {Ghosh}, {Jensen}, {Kazemi}, {Koopmans},
  {Labropoulos}, {Mevius}, {Martinez}, {Mellema}, {Offringa}, {Pandey},
  {Schaye}, {Thomas}, {Vedantham}, {Veligatla}, {Wijnholds}, \&
  {Yatawatta}}]{patil14}
{Patil}, A.~H., {Zaroubi}, S., {Chapman}, E., {Jeli{\'c}}, V., {Harker}, G.,
  {Abdalla}, F.~B., {Asad}, K.~M.~B., {Bernardi}, G., {Brentjens}, M.~A., {de
  Bruyn}, A.~G., {Bus}, S., {Ciardi}, B., {Daiboo}, S., {Fernandez}, E.~R.,
  {Ghosh}, A., {Jensen}, H., {Kazemi}, S., {Koopmans}, L.~V.~E., {Labropoulos},
  P., {Mevius}, M., {Martinez}, O., {Mellema}, G., {Offringa}, A.~R., {Pandey},
  V.~N., {Schaye}, J., {Thomas}, R.~M., {Vedantham}, H.~K., {Veligatla}, V.,
  {Wijnholds}, S.~J., \& {Yatawatta}, S. 2014, ArXiv e-prints

\bibitem[{{Petrovic} \& {Oh}(2011)}]{petrovic11}
{Petrovic}, N. \& {Oh}, S.~P. 2011, MNRAS, 413, 2103

\bibitem[{{Planck Collaboration}(2013)}]{gouldbelt}
{Planck Collaboration}. 2013, A\&A, 557, A53

\bibitem[{Pober {et~al.}(2013)Pober, Parsons, Aguirre, Ali, Bradley, Carilli,
  DeBoer, Dexter, Gugliucci, Jacobs, Klima, MacMahon, Manley, Moore, Stefan, \&
  Walbrugh}]{pober13}
Pober, J.~C., Parsons, A.~R., Aguirre, J.~E., Ali, Z., Bradley, R.~F., Carilli,
  C.~L., DeBoer, D., Dexter, M., Gugliucci, N.~E., Jacobs, D.~C., Klima, P.~J.,
  MacMahon, D., Manley, J., Moore, D.~F., Stefan, I.~I., \& Walbrugh, W.~P.
  2013, The Astrophysical Journal Letters, 768, L36

\bibitem[{{Ricciardi} {et~al.}(2010){Ricciardi}, {Bonaldi}, {Natoli},
  {Polenta}, {Baccigalupi}, {Salerno}, {Kayabol}, {Bedini}, \& {de
  Zotti}}]{ricciardi2010}
{Ricciardi}, S., {Bonaldi}, A., {Natoli}, P., {Polenta}, G., {Baccigalupi}, C.,
  {Salerno}, E., {Kayabol}, K., {Bedini}, L., \& {de Zotti}, G. 2010, MNRAS,
  406, 1644

\bibitem[{Santos {et~al.}(2005)Santos, Cooray, \& Knox}]{santos05}
Santos, M.~G., Cooray, A., \& Knox, L. 2005, ApJ, 625, 575

\bibitem[{Santos {et~al.}(2010)Santos, Ferramacho, Silva, Amblard, \&
  Cooray}]{santos10}
Santos, M.~G., Ferramacho, L., Silva, M.~B., Amblard, A., \& Cooray, A. 2010,
  MNRAS, 406, 2421

\bibitem[{Thompson {et~al.}(2001)Thompson, Moran, \& Swenson}]{thompson01}
Thompson, A.~R., Moran, J.~M., \& Swenson, Jr., G.~W. 2001, {\em Interferometry
  and Synthesis in Radio Astronomy} (John Wiley and Sons)

\bibitem[{Trott {et~al.}(2012)Trott, Wayth, \& Tingay}]{trott12}
Trott, C.~M., Wayth, R.~B., \& Tingay, S.~J. 2012, The Astrophysical Journal,
  757, 101

\bibitem[{Vedantham {et~al.}(2012)Vedantham, Shankar, \&
  Subrahmanyan}]{vedantham12}
Vedantham, H., Shankar, N.~U., \& Subrahmanyan, R. 2012, The Astrophysical
  Journal, 745, 176

\bibitem[{Wang {et~al.}(2006)Wang, M.Tegmark, Santos, \& Knox}]{wang06}
Wang, X., M.Tegmark, Santos, M.~G., \& Knox, L. 2006, ApJ, 650, 529

\bibitem[{Zaldarriaga {et~al.}(2004)Zaldarriaga, Furlanetto, \&
  Hernquist}]{zal04}
Zaldarriaga, M., Furlanetto, S.~R., \& Hernquist, L. 2004, ApJ, 608, 622

\bibitem[{Zibulevsky \& Pearlmutter(2001)}]{zibulevsky01}
Zibulevsky, M. \& Pearlmutter, B.~A. 2001, Neural Computation, 13, 863

\end{thebibliography}

\end{document}